# Detecting disease progression from animal movement using hidden Markov models

Dongmin Kim[1,2*], Théo Michelot[3], Katherine Mertes[4], Jared A. Stabach[4], John Fieberg[5]

[1.] *Department of Ecology, Evolution, and Behavior, University of Minnesota, St. Paul, MN, USA*

[2.] *Department of Organismic and Evolutionary Biology, Harvard University, Cambridge, MA, USA*

[3.] *Department of Mathematics and Statistics, Dalhousie University, Halifax, Nova Scotia, Canada*

[4.] *Conservation Ecology Center, Smithsonian's National Zoo & Conservation Biology Institute, Front Royal, VA, USA*

[5.] *Department of Fisheries, Wildlife and Conservation Biology, University of Minnesota, St. Paul, MN, USA*

## Abstract

1. An understanding of disease dynamics is important for managing wildlife populations and for quantifying the potential risk of spillover to domestic animals and humans, yet it is difficult to collect data on the infection status of wild, free-ranging animals. Pathogen and parasite infections alter host movement behavior, suggesting that it may be possible to infer infection status from observations of animal movement.

2. We propose a hidden Markov model (HMM) framework where an unobserved state process infers an animal's infection status from its observed behaviors, thus linking movement trajectories to epidemiological processes. This approach is consistent with compartmental models in epidemiology, where individuals may transition among states such as "susceptible", "infected", "recovered", and "dead", and formally connects observed animal movement parameters to disease dynamics.

3. We compiled movement data from 84 reintroduced scimitar-horned oryx (*Oryx dammah*), of which 38 were confirmed dead in the field and 6 were sampled for disease testing. We demonstrate several model formulations to show how HMMs can be tailored to epidemiological assumptions, including (1) constraints on transition probabilities (to preclude or include recovery), (2) covariate effects (to investigate the influence of factors that may affect disease transmission), and (3) hierarchically structured HMMs (HHMMs; to capture state transitions at multiple scales). We compared veterinary diagnostic reports to model outputs and found that HMMs with constrained transition probabilities successfully identified infection-associated reductions in movement, whereas unstructured models did not accurately detect disease progression. We also simulated movement data for individuals that could recover from an infected state, and found that constrained HMMs also accurately classified susceptible, infected, and recovered states.

4. By demonstrating the flexibility of (H)HMMs in capturing different disease scenarios, and a workflow for appropriate model selection, we provide a transferable workflow for detecting infection from animal movement data.

5. Our approach has the potential to improve wildlife disease surveillance, inform management of vulnerable populations, and enhance understanding of disease dynamics.

**Keywords:** Hidden Markov model, Wildlife disease, Disease ecology, Movement ecology, Conservation monitoring, Animal telemetry, Reintroduction

# 1. Introduction

Infectious diseases are an integral part of the life history of animals, influencing host movement and resulting in increased rates of morbidity and mortality (May & Anderson, 1978; Holmes et al., 1972). Over the past decade, classic epidemiological models such as the Susceptible, Infected, and Recovered (SIR) framework, along with laboratory experiments, have provided important insights into how infection alters host activity and movement (Hall et al., 2022). For instance, theoretical models and experimental studies show that many infected hosts decrease their activity compared to non-infected hosts (Binning et al., 2017; Bradley and Altizer, 2005; Debeffe et al., 2014; Goodman and Johnson, 2011; Kim & Shaw, 2021; Poulin, 1994; Oppliger et al., 1996). In controlled experiments, different host taxa have been shown to decrease their movement capacity (i.e., distance traveled, time spent moving) after infection (Poulin, 1994). Similarly, mathematical models predict that infection heterogeneity (i.e., variable infection intensity and cost) can lead to partial migration, where infected hosts with higher infection costs evolve not to migrate (Balstad et al., 2020).

Despite growing evidence from theoretical and experimental studies of disease-related changes in animal movement, few empirical examples directly link disease status to movement (Binning et al., 2017; Chretien et al., 2023). Debeffe et al. (2014) showed that nematode abundance decreased roe deer (*Capreolus capreolus*) body condition and dispersal propensity. Similarly, Dekelaita et al. (2023) found that desert bighorn sheep (*Ovis canadensis nelsoni*) infected with pneumonia had lower mean daily movement rates and were significantly less likely to make intermountain movements. Similarly, Barrile et al. (2024) showed that mule deer (*Odocoileus hemionus*) infected with chronic wasting disease (CWD) moved more slowly than non-infected individuals. In contrast, Spaan et al. (2019) found no direct effects of infection by gastrointestinal parasites or microparasites on dispersal in adult female Cape buffalo (*Syncerus caffer*). These varied results emphasize the need for further research to investigate the complex relationships between infectious diseases and movement in free-roaming animals.

Recent outbreaks of the highly pathogenic avian influenza (HPAI) H5N1 viruses, carried by domestic and wild birds, have spread to other wildlife, domestic mammals, and humans (Caserta et al., 2024; Lambertucci et al., 2025; Garg et al., 2025). Transmission of these viruses has resulted in the death of thousands of domestic and wild animals, and a human death from the H5N1 infection has recently been reported (Kang et al., 2024; Plaza et al., 2024; Uyeki et al., 2024). Severe acute respiratory syndrome coronavirus 2 (SARS-CoV-2), which causes COVID-19 in humans, has also been detected in wildlife, such as large ungulates throughout North America (Pickering et al., 2022; Hewitt et al., 2024). It is unclear how these viruses are transmitted and persist in ungulate populations, or whether they can be transmitted between ungulates and humans (Pepin et al., 2017; Wilber et al., 2022). To prevent large or zoonotic disease outbreaks, we need better tools for detecting disease and quantifying disease progression.

The progressive miniaturization of electronics has made it possible to track the detailed movements of various species over larger spatial extents and longer time periods (Nathan et al., 2022). Presumably, if infected animals move differently from non-infected animals, movement data could be used to detect and infer disease dynamics. However, contemporaneously assessing the disease status of tracked animals for specific pathogens remains a persistent challenge. Recapturing or otherwise obtaining biological samples from a sufficient number of tagged animals is extremely expensive and time-consuming, as are the rigorous laboratory methods (e.g., molecular techniques for blood/tissue samples and assessing leukocyte count) required to infer infection status (Chretien et al., 2023). These persistent challenges have long limited the development, parameterization, and evaluattion of movement-based models of disease progression.

In epidemiology, biostatisticians often use hidden Markov models (HMMs) to detect disease progression in patients from surveillance and hospital infection data (Green & Richardson, 2002; Cooper & Lipsitch, 2004; Watkins et al., 2009; Li et al., 2021). HMMs are also well established in movement ecology, where they are used to infer changes in animal behavior from tracking data. For example, HMMs have been used to identify foraging and exploring behaviors (Morales et al., 2014; McClintock and Michelot, 2018) and infer survival or mortality (Runde et al., 2020). Despite their widespread uses, we are unaware of any applications of HMMs to detect disease progression in wildlife using movement data.

Here, we fit multiple HMM structures to both tracking data from free-ranging scimitar-horned oryx (*Oryx dammah*, hereafter "oryx") and simulated movement data, using veterinary diagnostics and *in situ* observations for oryx, and known states for simulated trajectories, to evaluate predictions. To illustrate the range and flexibility of HMMs, we develop three model structures that reflect different assumptions about host-disease dynamics: (1) an SIS model, where individuals transition between susceptible and infected states and remain vulnerable to reinfection; (2) an SI model, where individuals are unlikely to return to a susceptible state after infection ; and (3) a SIR model, where individuals can recover and gain immunity to future infections. We use additional simulations to demonstrate how analysts can generate movement tracks and disease states tailored to different epidemiological scenarios. Finally, we show how each model structure  provides generalizable insights into the links between movement and disease infection.

## 2. Materials and Methods

### 2.1. Study System

Oryx are large African antelope native to the seasonal grasslands fringing the Sahara Desert. Once widespread across West, Central, and North Africa (Brouin, 1950; Malbrant, 1952; Gillet, 1965, 1969; Newby, 1988), increased hunting pressure – due to expanded access to modern weaponry and 4x4 vehicles across the Sahel – as well as habitat degradation and increasing competition with domestic livestock, led to the species' decline (Gillet, 1965, 1969; Newby, 1978a, 1978b, 1988; Dragesco-Joffé, 1993). Oryx were last reported in the wild in 1988 (Beudels-Jamar et al., 1998) and were classified as Extinct in the Wild by the IUCN in 1999 (East, 1999). Due to ongoing reintroduction efforts , oryx were reclassified to Endangered by the IUCN in 2023 (Wallis, 2023).

## 2.2. Oryx reintroduction methods

Since 2013, the Environment Agency – Abu Dhabi (EAD), in partnership with the government of Chad and Sahara Conservation and supported by other technical partners, has led a large reintroduction initiative based in the Réserve de Faune de Ouadi Rimé-Ouadi Achim (RFOROA) in central Chad (Chuven et al., 2018). Beginning in 2016, groups of 19-73 oryx were translocated from the World Herd managed in Abu Dhabi and released into the reserve. Oryx were translocated during cooler months (ca. October - March in Abu Dhabi and Chad) and transported to the RFOROA by cargo plane and heavy truck. Oryx were held in 24-ha enclosures for 2-8 months to adjust to local environmental conditions (Appendix S1). During this acclimation period, oryx were gradually transitioned to a natural diet by decreasing the amount of hay, pellets, and water provided. The majority (>95%) of reintroduced oryx were fitted with GPS collars (Vectronic Aerospace GMBH; Berlin, Germany) programmed to collect positions every 1-4 hours. Oryx were collared during brief periods of restraint (< 10 min) in a drop-chute device (Fauna TAMER Jr; Fauna Research Inc., Red Hook, New York, USA) 1-2 weeks before release (Appendix S1). On the release date, a gate was opened and animals passively exited the enclosure. Animal handling methods were approved by the Animal Care and Use Committee (ACUC) at the Smithsonian's National Zoo and Conservation Biology Institute (NZCBI) and authorized under a cooperative agreement between Sahara Conservation and the Chadian Ministère de l'Environnement, de la Pêche, et du Développement Durable (NZP-IACUCs #15-32, 17-21, 18-38, and SI-23051). The reintroduced oryx population in the RFOROA is currently estimated to contain 575 ± 348 animals (Wacher et al., 2023).

## 2.3. Mass mortality event

From late August to early November 2018, a massive mortality event (MME) occurred in the reintroduced oryx population due to several pathogenic, bacterial, and parasitic infections, including Rift Valley Fever (RVF), Peste des Petits Ruminants (PPR), and Babesiosis (Appendix S1). Anomalously high rainfall during the 2018 rainy season (ca. twice the typical cumulative rainfall for July) was linked to rapid, extreme increases in of biting insects, creating favorable conditions for RVF and other vector-transmitted diseases, as well as bacterial and parasitic infections (Chardonnet, 2019). Thirty-eight oryx mortalities were recorded between August and November 2018, with mortalities distributed evenly across sex and age classes. More than half of the oryx released during the 2018 rainy season died within two months. This outcome stands in stark contrast to both previous (2016-2017) and successive years (2019-2023), when more than 80% of each release group survived to one year after release. Based on findings from a contemporaneous veterinary field mission (Chardonnet et al. 2019), subsequent disease tests at a reference laboratory, and the characteristic bell-shaped distribution of oryx mortalities over time, the elevated mortality rate during this period was attributed to disease.

A veterinary field mission during the MME collected blood, smear, swab, and tissue samples from ten oryx: five rom examinations of live animals released before 2018 that appeared sick and five from field necropsies of animals released during the MME period. These samples were subsequently tested by the Institut de Recherche en Elevage pour le Développement (IRED; N'Djamena, Chad) and the Centre de Coopération Internationale en Recherche Agronomique pour le Développement (CIRAD; Montpellier, France). All sampled oryx tested positive for various infectious diseases (Chardonnet, 2019), including parasitic, bacterial, and pathogenic infections (i.e., co-infection), and all but one died during the MME . Three individuals tested positive for parasitic diseases (e.g., Babesiosis), seven individuals tested positive for bacterial infections (e.g., Pasteurellosis), five individuals tested positive for RVF, and six individuals tested positive for PPR (Chardonnet, 2019).

## 2.4. Study period and study population

Reintroduced oryx in the RFOROA experience three seasons: a hot dry season (March 13 - July 10), a short rainy season (July 11 - October 1), and a longer cool dry season (October 2 - March 12; Whyle et al. 2025). We constrained our study period to July 11 - October 1, 2018 , to focus on the massive mortality event and limit the influence of seasonality on oryx movement behavior. Of 117 oryx tracked during this period, we included 84 in our study: 46 oryx that lived through the entire study period and 38 oryx that died during the study period, 6 of which tested positive for RVF and other co-infections (Appendix S1; Figure S1). We removed 13 individuals with coarser GPS fix rates and 20 individuals for which fewer than 100 locations were recorded, either due to mortality or tracking device malfunction. Of these 84 oryx, 67 were released during the 2018 rainy season , and17 were released in 2017 ($n = 5$ released in January 2017 and $n = 12$ released in August 2017). After data cleaning, six of the ten oryx sampled during the MME period had more than 100 hourly locations and were included in our final data set. Four of the six sampled oryx tested positive for RVF and other diseases like Babesiosis and PPR (Appendix S1) and exhibited high viral loads of RVF; and two of these oryx also lacked antibodies for RVF, indicating that its onset and development progressed so rapidly that infected animals had not had sufficient time to produce antibodies. Five of the six sampled oryx tested positive for other diseases, such as PPR and Baesiosis (Appendix S1; Chardonnet, 2019). Our final movement dataset contained 111,158 locations from 84 individuals (Appendix S1).

## 2.5. Animal movement data and environmental covariates

Given that oryx experienced a relatively long acclimation period before release, we did not expect animals to exhibit a post-release handling response. However, as a conservative approach to remove potential effects of release-related stress, we excluded data during the 24 hours immediately after release, per Northrup et al. (2014). We then visually explored mean daily step length versus time since release for all oryx in our final dataset (up to 90 days) to check for transitory dynamics as animals became familiar with a novel environment (Appendix S1; Figure S2).

All reintroduced oryx were regularly observed by an ecological monitoring team based in the RFOROA, including assessments of animal body condition. Body condition was scored using a 9-point scale developed in a managed care setting (Eyres et al., 2019): 1-3 was considered "underweight", 4-6 was considered "optimal", and 7-9 was considered "overweight". During the MME, recently reintroduced animals received increased monitoring attention and were directly observed multiple times each week. We calculated weekly mean body condition scores for each individual in our final data set for which at least two weeks of body condition scores were collected.

In the highly seasonal, precipitation-limited grasslands and savannas of the RFOROA, herbaceous vegetation exhibits strong annual cycles of green-up and senescence, while trees and shrubs maintain some photosynthetic activity across seasons. We thus derived a shrub cover covariate by summing 14-day mean MODIS NDVI (250m resolution) measurements across three years, centered on the MME period, in Google Earth Engine (Gorelick et al., 2017). We also calculated the time since release for each oryx in our final dataset, and included this as a covariate to allow for transitory dynamics and to test whether oryx exhibit more exploratory movements shortly after being released in a new environment.

## 2.6. Hidden Markov model (HMM) formulations

We used the momentuHMM package in R (McClintock and Michelot, 2018) to fit a series of HMMs to the oryx movement data to identify infection status (e.g., susceptible, infected, recovered, and dead) and behavioral state (e.g., exploring, resting) within each infection status. . We included the covariates shrub cover and time since release to investigate how these factors affect oryx movements and interact with disease and behavior dynamics . Based on the unprecedented mortality rate of oryx released during the 2018 rainy season and the extremely low number of observed recoveries (n=1), we treated death as an absorbing state (i.e., transition probabilities out of a "death" state were fixed to zero).

We defined state-dependent distributions for two movement variables, representing step lengths ($sl_t$) and turning angles ($ta_t$) at time $t$. We assumed step lengths ($sl_t$) followed a gamma distribution with state-specific shape ($k_b$) and scale ($\theta_b$) parameters, and that turning angles ($ta_t$) followed a von Mises distribution with state-specific mean ($\mu_b$) and concentration parameters ($\phi_b$):

$$sl_t | (S_t = b) \sim gamma(k_{b,} \theta_b)$$

$$ta_t | (S_t = b) \sim von\ Mises(\mu_{b,} \phi_b)$$

for each state $S_t$. To avoid converging to a local rather than global maximum, we considered 15 different starting values when fitting each HMM and selected final starting values from the model with the lowest AIC (Appendix S2).

## 2.7. Choosing the number of HHM states

The choice of the number of HMM states is challenging because model selection criteria tend to favour complex models with low interpretability, and it is usually preferable to make a decision based on domain expertise (Pohle et al., 2017). We initially considered HMMs with 3, 4, or 5 states (Appendix S2). In the simplest model (3-state), we tentatively interpreted the states as susceptible, infected, and dead, similar to a common structure for SI models. In the more complex models, we sought to split susceptible (4-state) or susceptible and infected (5-state) phases into two behavioural states (resting and exploring).

From initial comparisons of 3, 4, and 5-state HMMs, we found that the 3-state model could not differentiate between larger and smaller movements (Appendix S2). On the other hand, the additional flexibility of splitting "infected" status into two states in the 5-state HMM led the model to produce implausible classifications, with very long periods of infection(Appendix S2). We thus selected a 4-state HMM as most appropriate for our oryx MME, and focus on various implementations of 4-state models.

## 2.8. Unconstrained 4-state HMMs: Susceptible-Infected-Susceptible (SIS)

We began by fitting an unconstrained HMM with four states, which were intended to capture: (1) susceptible exploring (i.e., exploration movements by a susceptible but uninfected animal; SE), (2) susceptible resting (SR), (3) infected (I), and (4) death (D) (Figure 1a; Figure 2a). The "exploring" and "resting" states denote periods of faster, longer vs slower, shorter movements. We expected the latter state to capture a mixture of low-activity behavioral states, such as ruminating, in addition to resting. We modeled transition probabilities as a function of shrub cover and time since release using a multinomial logistic formulation - except for the transitions between resting and death states, which were assumed to have constant probability (Patterson et al., 2009). This model structure predicted frequent transitions between "infected" and "susceptible" states, suggesting that animals became infected and recovered many times each day - a biologically implausible sequence of events given the observed outcomes during the oryx MME. In the following sections, we use information from the oryx MME to derive realistic constraints for transition probabilities, and explore alternate HMM structures to represent alternative epidemiological scenarios.

## 2.9. Constrained 4-state HMMs: Susceptible-Infected (SI)

To capture a low probability of recovery, as observed in the 2018 oryx MME, we fit a 4-state HMM where transition probabilities from the infected state to susceptible states were fixed to zero (Figure 1b; Figure 2b). Reintroduced oryx frequently rest in tightly clumped groups under the shade offered by sparse trees and shrubs (i.e., in very close contact with potentially infected individuals). We thus further constrained the model such that non-infected oryx could only become infected when resting (via transition rate $\gamma_{24}$). We also assumed that transitions to the death state only occurred from either the susceptible resting state ($\gamma_{24}$) or the infected state ($\gamma_{34}$).

## 2.10 Constrained 4-state HMMs: Susceptible-Infected-Recovered (SIR)

Although recovery was unlikely in our oryx study system, individuals are often capable of recovering from a diseased state. We thus fit a 4-state SIR HMM, modeling transitions among susceptible, infected, recovered, and dead states (Figure 1c; Figure 2d). To simplify model structure, we assumed that once individuals recovered, they remained in the recovered state, while other assumptions followed the constrained model described in 2.9. However, a constrained SIR HMM applied to the oryx movement data could not distinguish between susceptible and recovered states, because no oryx exhibited extended periods of large steps, then small steps, and subsequently large steps again, during the MME period (Appendix S2). To address this limitation, we simulated daily movement paths that varied with infection status including recovery, and then fitted the constrained SIR 4-state HMM to illustrate how a recovery state can be incorporated (see Section 2.12 for details).

## 2.11. 4-state hierarchical hidden Markov model (HHMM)

We next fitted a hierarchical hidden Markov model (HHMM) with two temporal scales (Leos-Barajas et al., 2017a): (1) a coarse scale allowing transitions between "susceptible", "infected", and "dead" states to occur at 3-day intervals, and (2) a fine scale allowing susceptible individuals to transition between two behavioral states (interpreted as exploring and resting; Figure S1) at each hourly movement step (Figure 2c). Infected and dead individuals were assumed to have only a single fine-scale behavioral state. We modeled state-dependent step-length and turn-angle distributions at the fine scale, and no data streams were connected to the coarse scale. We fitted both constrained and unconstrained HHMMs.

For the constrained HHMM, we fixed transition probabilities at the coarse scale to preclude recovery, similar to our constrained 4-state HMM. Within the susceptible coarse-scale state, we modeled fine-scale transition probabilities as a function of shrub cover and time since release. We specified starting values for estimating parameters of the state-dependent distributions and transition probabilities based on estimates from the constrained 4-state HMM, . In the unconstrained HHMM, transitions between Infected and Susceptible were allowed every 3 days (Appendix S2), whereas in the unconstrained HMM all transitions could occur hourly.

## 2.12. Simulation framework

We simulated 30 days of hourly movement trajectories for 20 individuals. First, we generated daily infection-related behavioral states using a first-order Markov process with the transition probability matrix described in Figure 1c. Based on each daily infection status, we then generated 24-hour movement paths, using four distinct state-specific redistribution kernels, empirically derived from the oryx movement data, representing different space use patterns per disease status (susceptible, infected, recovered, or dead). Susceptible and recovered individuals were assumed to exhibit more exploratory movement, infected individuals reduced movement, and dead individuals no movement. Locations from dead individuals were randomly perturbed within 15m to represent potential GPS error by a stationary collar. We set the transition probability between "Recovered" and "Infected" states to 0 to represent the immunity of recovered individuals to future infection.

## 2.13 Model validation

For the oryx data, we overlaid estimated state-dependent distributions on empirical distributions of step lengths and turn angles (Figures 3 and 4). We also plotted time series of mean daily step lengths for all individuals, with a focus on the six individuals that were tested by veterinarians (Figures 5 and 6). We also used other diagnostic tools, such as pseudo-residual and ACF plots, to check model assumptions (Appendix S2). Lastly, we evaluated the models' ability to correctly classify the 38 individuals that were confirmed dead. We initially included known individual fates in a semi-supervised approach, but these models did not converge (Appendix

S2). For the simulated movement trajectories, we evaluated model performance by comparing Viterbi-inferred states to known states using common performance metrics, precision, accuracy, recall, and F1 scores (Appendix S2).

# 3. Results

## 3.1. Unconstrained (H)HMMs (SIS model)

We first focus on results pertaining to the interpretation of the states in the unconstrained 4-state HMM, to highlight flaws of this model formulation. In the unconstrained 4-state HMM, transition probabilities between the tentative "Susceptible" (Exploring or Resting) and "Infected" states were large, and the most likely state sequence predicted by the Viterbi algorithm displayed frequent transitions between them. Overall, the median number of transitions for an individual between the "Susceptible" states and the "Infected" state was 79 in each direction. In addition, the mean step length in the third state (tentatively "Infected") was 4 m, which is less than the ca. 10m spatial error of the GPS collar fit to reintroduced oryx, indicating virtually no movement at all (Appendix S2). These results indicate that the states of the unconstrained 4-state HMM did not match our intended interpretation.

The unconstrained HHMM also exhibited unrealistic recoveries (i.e., transitions from Infected to Susceptible), but these occurred less frequently than in the unconstrained 4-state HMM (Figure 5; Appendix S2). All 84 oryx exhibited transitions from Infected to Susceptible in the unconstrained 4-state HMM (Figure 5a), whereas only 31 of the 84 oryx transitioned from Infected to Susceptible in the unconstrained HHMM (Appendix S2, Figure S5). In the unconstrained HHMM, the median number of transitions per individual from the "Susceptible" to the "Infected" state was 1 (max = 4, number of individuals with more than 1 transition = 21), whereas the median for the opposite direction – from "Infected" to "Susceptible" – was 0 (max = 4, number of individuals with more than 1 transition = 11).

## 3.2 Constrained (H)HMMs (SI model)

### 3.2.1 State-dependent distributions

The estimated state-dependent distributions from the constrained 4-state HMM largely matched our *a priori* expectations for the interpretation of the four states (Figure 3; Appendix S2). The Susceptible Exploring state had long step lengths and small turning angles, corresponding to fast, directed movement. The Susceptible Resting and Infected states had shorter step lengths and flat distributions of turning angles (i.e., undirected movement). Finally, the Dead state had the shortest step lengths and a turning angle distribution centered on $\pi$, likely an artifact of measurement error (Hurford, 2009). The constrained (and unconstrained) HHMMs had state-dependent distributions that were very similar to the constrained HMM (Figure 4; Appendix S2).

### 3.2.2 State decoding and covariate effects

The constrained 4-state HMM and constrained HHMM captured the large overall decline in movement that occurred in many of the individuals' time series (Figure 5) and attributed these changes to transitions from Susceptible to Infected states. The constrained 4-state HMM correctly classified 33 of the 38 confirmed dead individuals, whereas the constrained 4-state HHMM correctly classified 27 of the 38 individuals (Appendix S2). We further compared the estimated state sequences between the constrained HMM and the constrained HHMM for the six dead individuals that tested positive for pathogen, bacterial, and parasitic diseases. Four out of the six individuals exhibited "underweight" body condition scores at the end of their movement trajectories (Figure 6). Although the estimated state sequences for the constrained HMM and the constrained HHMM were similar in most cases, there were some notable differences that relate to the inferred timing of infection. For example, the HMM (Figure 6a) predicted that individual 80 died after the infection, while the HHMM (Figure 6b) predicted that the individual suddenly died without getting infected. Both models failed to detect the death of individual 49, and predicted that this individual was infected and alive at the end of its trajectory.

We did not detect a clear effect of shrub cover or time since release on transition probabilities from Susceptible Exploring to Susceptible Resting and from Susceptible Resting to Infected in any of the models (Appendix S2).

### 3.3 Constrained HMMs (SIR model)

The constrained 4-state SIR HMM applied to the oryx data could not differentiate Susceptible and Recovered states (Appendix S2), highlighting that this case study was not suitable for testing whether HMMs can classify recovery. When applied to simulated data, the 4-state SIR HMM successfully recovered the true states, achieving F1 scores of 0.996 for Susceptible, 0.987 for Infected, 0.999 for Recovered, and 0.991 for Dead states (Appendix S2; Appendix S2).

# 4. Discussion

In this study, we demonstrate that HMMs with structures, state numbers, and transition probabilities parameterized based on an observed disease scenario accurately predicted disease progression and fate from animal movement data. Specifically, based on disease testing results from our study system, we constrained the transition probability matrix so that infected individuals could not recover (Figure 1b), which improved model performance over unconstrained HMMs (Figure 6; Appendix S2). We also highlighted how analysts may structure HMMs to reflect different epidemiological dynamics, such as (1) an SIS model, where individuals can transition back and forth between infected and susceptible; (2) an SI model, where transitions from infected to susceptible are unlikely; and (3) a SIR model, where individuals may recover from, and then remain immune to, a disease of interest. We encourage analysts to formulate HMMs based on available data and biological knowledge of their own study systems. Importantly, this approach requires that diseased individuals exhibit changes in their movement behavior . However, additional data streams - for example, tri-axial accelerometry data often collected by animal tracking devices - may enable the detection of diseases with less extreme effects on animal movement, or subclinical infections.

## 4.1. Biological assumptions, model interpretation, and practical limitations

Previous studies have found that recovery rates for RVF and PPR alone are relatively high (~65% in wildlife and ~90% in livestock) and that recovered individuals have lifelong immunity (Jost et al., 2010; Hartman, 2017). However, six of the oryx sampled during our study period tested positive for several diseases, including pathogenic (RVF and PPR), but also parasitic (Babesiosis) and bacterial (Pasteurellosis) infections (Appendix S1). Also, almost half of the study population ($n$ = 38) died shortly after release (less than a month; Appendix 1). This suggests that an infected oryx's condition was likely to worsen until it died, or that it might slowly recover, but over a time frame longer than the study period. Based on these test results and *in situ* observations, we thus specified a model formulation in which recoveries were unlikely over the study period and multiple recoveries and reinfections were virtually impossible. Additional testing data and detailed biological information about pathogen effects on scimitar-horned oryx would be needed to refine these epidemiological assumptions.

We constructed (H)HMMs with several behavioral states based on *a priori* expectations about movement behavior by large terrestrial herbivores and the typical progression of infectious diseases (Adam et al., 2019). Our results support a hypothesis posed by previous studies that diseases reduce the movement and activity of large herbivores and increase resting behavior (Barrile et al., 2024; Debeffe et al., 2014; Morelle et al., 2023). However, it is critical to inspect estimated parameters in detail to aid in model interpretation and ensure that state labels align with biologically meaningful behavioral states when using any form of unsupervised classification. If the infectious or behavioral state of an animal is known during part of the study period (e.g., based on direct observations), this information can be passed to the model in a semi-supervised approach (Leos-Barajas et al., 2017b).

The unconstrained 4-state HMM (SIS) predicted frequent switches between all states within a day. These biologically implausible predictions indicate that the states of this HMM did not correspond to the epidemiological states of interest, and suggest that this model formulation was not appropriate. We additionally fitted a model that included a recovered state and found that the model had a difficult time differentiating susceptible and recovered states (Appendix S2). Ultimately, we opted for a constrained HMM where recoveries were precluded over the study period; we found that the resulting parameter estimates and predicted state sequences matched our expectations and the biological evidence observed in the field, including the high mortality rate, high viral load, and lack of antibodies in the tagged individuals (Figures 3, 4, 5, Appendix S1). This showcases the workflow that is required to identify the statistical assumptions that lead to an appropriate model for the data.

We also considered an unconstrained model with a hierarchical structure where transitions between disease states occurred at a coarser time scale (3 days) than the GPS telemetry observations (1 hour). While this approach worked well for some individuals, it predicted multiple recoveries for around one-fourth of the tagged animals, which was not consistent with observed disease dynamics (Appendix S2, Figure S5). We also compared the constrained 4-state HMM to a constrained HHMM (Figures 5 and 6) and found that some results (e.g., classification of dead individuals) appeared to favor the constrained 4-state HMM over the HHMM. With HHMMs, the scale of the coarse level transitions (in our case, 3 days) will influence how frequently these transitions can occur and should be chosen to approximate assumed disease dynamics. However, because the models are formulated in discrete time, it may be challenging to do so. With longer time intervals, it becomes more likely that individuals will transition well before the end of the time step, and thus, they will be in multiple states within the same time interval. Nonetheless, HHMMs may be beneficial with additional data sources observed at a different time resolution, like heart rate and body condition indices, aiding in infection tracking (Oliveira-Santos et al., 2021). For example, implanted accelerometer sensors can be used to measure the heart rates of tracked wildlife, which could serve as an indicator of infection status (i.e., the lower the heart rate, the greater the likelihood that an individual is infected; Leimgruber et al., 2023; Morelle et al., 2023). While we used weekly body condition scores for qualitative assessment, these scores could also prove useful as a state-dependent response variable if collected more frequently.

## 4.2. Future directions

We provided inferences from a population-level (H)HMM that assumed all parameters, state-dependent distributions, and state-changing dynamics were the same across individuals. Random effects can be included in HMMs to account for individual variability in transition probabilities or the parameters describing state-dependent distributions (McClintock, 2021; Michelot, 2025). Animals could then vary both in their activity budgets (e.g., time spent resting and exploring) and in their sensitivity to infection (Glennie et al., 2023). Future research could explore the performance of different types of HMMs (population-level, individual-level, and population-level with random effects) in inferring infection status from animal movement data. We also encourage users to adapt our simulation code to their own systems, and to further explore when these models can and cannot capture assumed disease dynamics.

We anticipate that HMMs may be useful for informing ongoing conservation activities by allowing protected area managers and researchers to evaluate whether managed populations or individuals are exhibiting disease-influenced movement behaviors, allowing field teams to provide preventative care. HMMs could also be used to estimate the timing of infection in managed populations, as illustrated in Figure 5, and link infection events to spatial landscapes and environmental predictors, and to observed contacts among individuals. Simulations from fitted models also could be used to gain insights into emerging theories regarding how infection influences movements, and thus, subsequent disease transmission dynamics. In summary, HMMs provide many new opportunities for disease and movement ecologists to address questions once limited to theoretical or experimental approaches. We encourage researchers to explore HMMs as a valuable method for understanding the ecological and evolutionary dynamics of disease-host interactions.

# Figure Captions

(a) Null transition probability matrix (SIS)

|     | SE | SR | I | D |
|-----|-----|-----|-----|-----|
| SE | $\gamma_{11}$ | $\gamma_{12}$ | $\gamma_{13}$ | $\gamma_{14}$ |
| SR | $\gamma_{21}$ | $\gamma_{22}$ | $\gamma_{23}$ | $\gamma_{24}$ |
| I | $\gamma_{31}$ | $\gamma_{32}$ | $\gamma_{33}$ | $\gamma_{34}$ |
| D | 0 | 0 | 0 | 1 |

(b) Study system transition probability matrix (SI)

|     | SE | SR | I | D |
|-----|-----|-----|-----|-----|
| SE | $\gamma_{11}$ | $\gamma_{12}$ | 0 | 0 |
| SR | $\gamma_{21}$ | $\gamma_{22}$ | 0 | $\gamma_{24}$ |
| I | 0 | 0 | $\gamma_{33}$ | $\gamma_{34}$ |
| D | 0 | 0 | 0 | 1 |

(c) Simulated system transition probability matrix (SIR)

|     | S | I | R | D |
|-----|-----|-----|-----|-----|
| S | $\gamma_{11}$ | $\gamma_{12}$ | 0 | $\gamma_{14}$ |
| I | 0 | $\gamma_{22}$ | $\gamma_{23}$ | $\gamma_{24}$ |
| R | 0 | 0 | $\gamma_{33}$ | $\gamma_{34}$ |
| D | 0 | 0 | 0 | 1 |

**Figure 1. (a).** Unconstrained transition probability matrix (SIS) where blue boxes represent transition probabilities between exploring and resting within the susceptible (S) state. Red boxes represent the transition probabilities from one disease state to another (infection or recovery). Elements in the green box indicate mortality rates. **(b).** We incorporated assumptions based on our observations from the field to construct our study-specific transition probability matrix (SI): (1) susceptible oryx only transitioned into an infected state when resting; (2) transitions to the death state only occurred from resting states; (3) infected individuals could not recover once infected; (4) individuals cannot leave the dead state (hence the zeros in the last row). **(c).** We generated simulated data with assumptions that individuals can recover from infection, and once they recover, they remain immune to further infection (SIR).

(a) Unconstrained 4-state HMM (SIS)

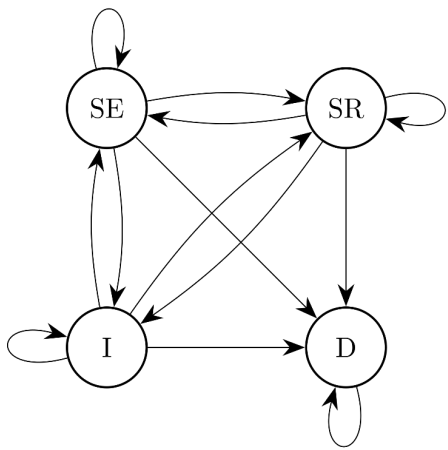

(b) Constrained 4-state HMM (SI)

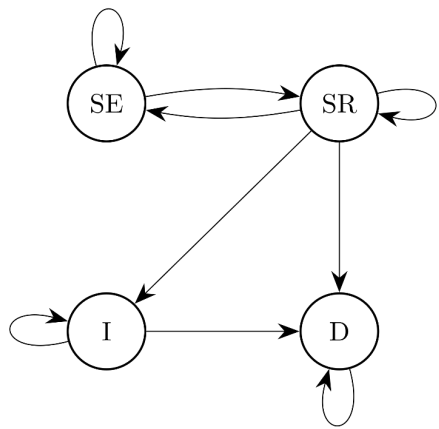

(c) Constrained 4-state HHMM (SI)

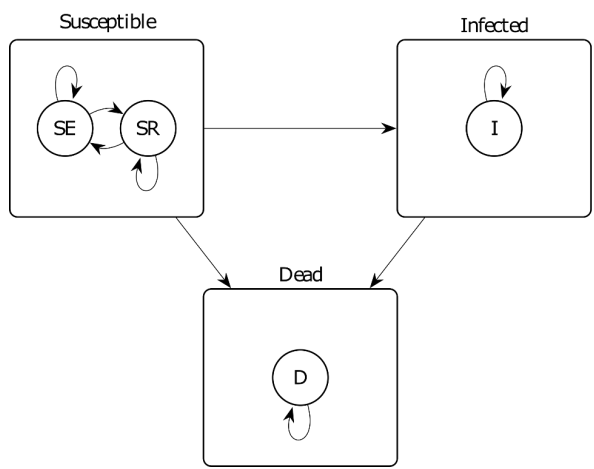

(d) Constrained 4-state HMM (SIR)

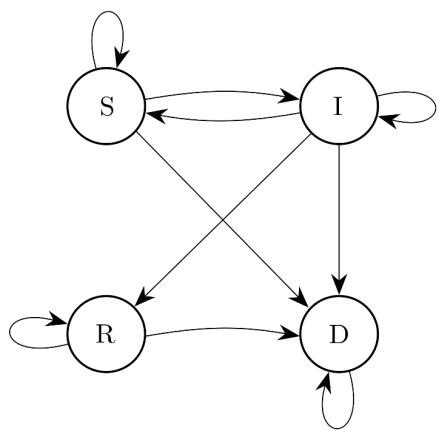

**Figure 2. (a).** Unconstrained 4-state (SIS) HMM, which allows transitions between all states other than the death state which is assumed to be an absorbing state, **(b).** Constrained 4-state (SI) HMM structure, **(c).** Constrained 4-state (SI) HHMM structure: the box indicates the coarse-level states (susceptible, infected, and dead) and transitions between circles within the boxes present the fine-level behavioral states, and **(d).** Constrained 4-state (SIR) HMM structure (susceptible, infected, recovered, and dead).

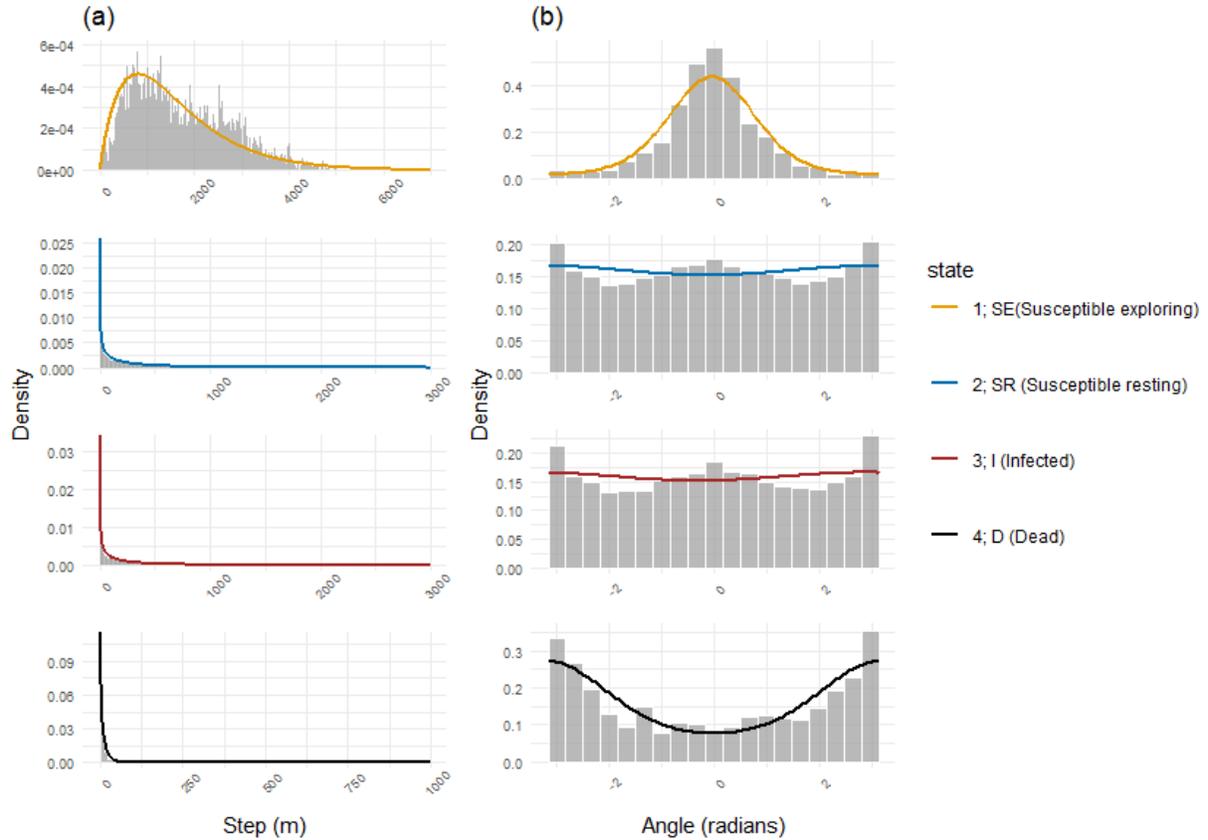

**Figure 3.** The constrained 4-state HMM: Estimated state-dependent distributions for **(a)** step lengths and **(b)** turn angles: Susceptible behavioral states ("1" - yellow and "2" - blue), infected behavioral state ("3" - brown), and death ("4" - black). Individuals took shorter steps when they were inferred to be infected ("3" - brown), and they took longer and more directed steps when exploring versus resting (states "1" - yellow and "2" - blue).

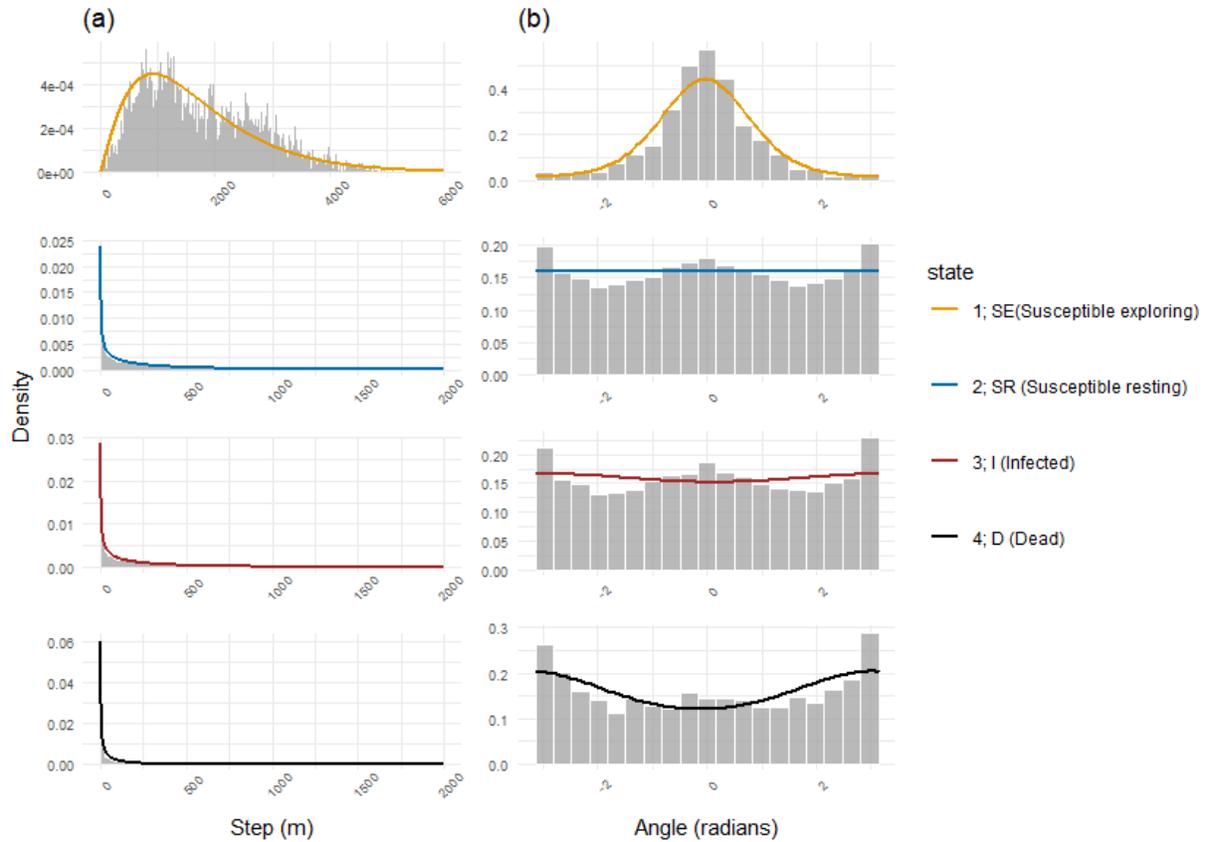

**Figure 4.** The constrained hierarchical hidden Markov model (HHMM): Estimated state-dependent distributions of **(a)** step lengths and **(b)** turn angles for Susceptible (exploring, "1" - yellow and resting, "2" - blue), Infected status (infected, "3" - brown) and Death status (death, "4" - black).

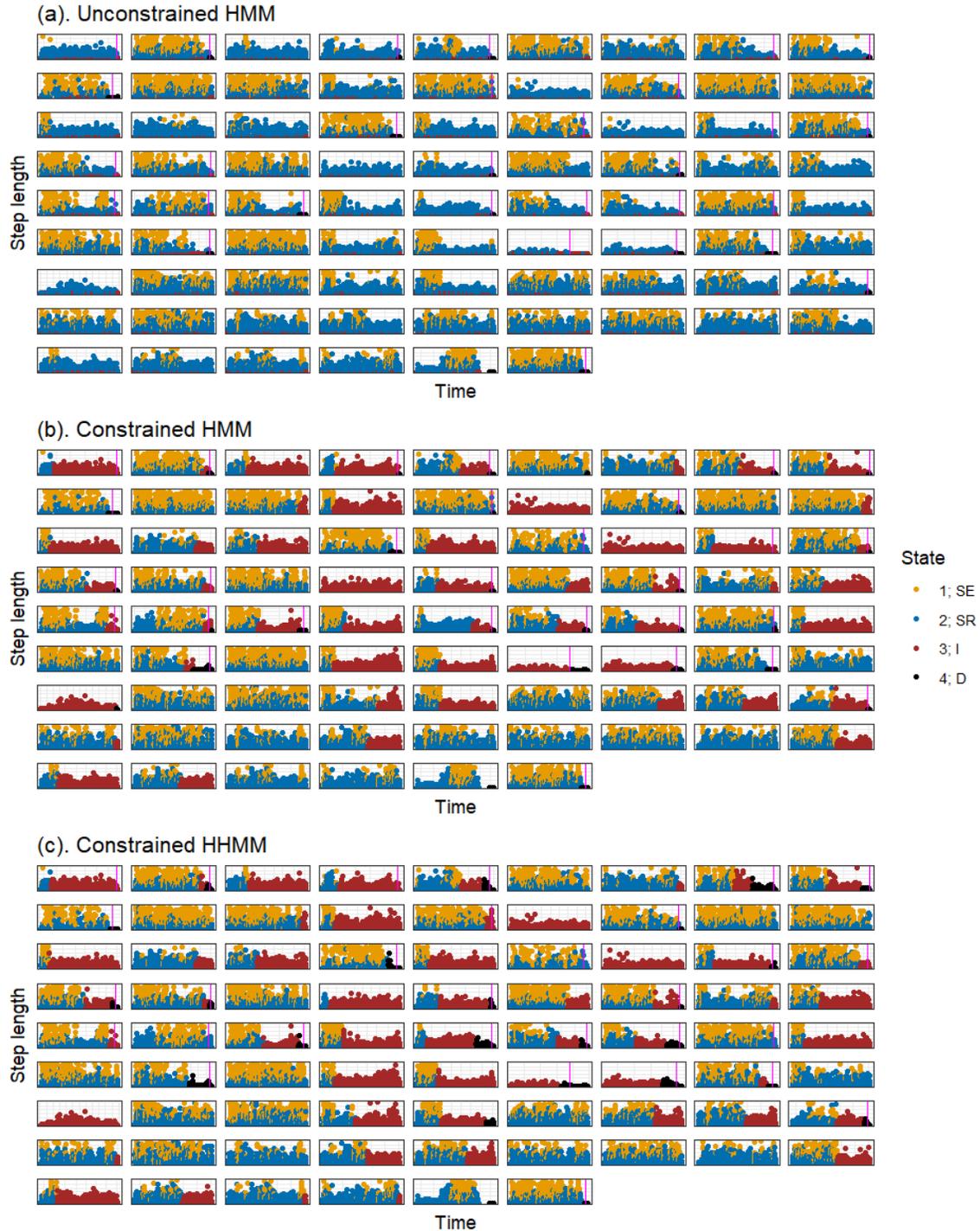

**Figure 5.** Time series of observed step lengths of oryx (n = 78) not tested in the field by veterinarians from (a) the 4-state unconstrained HMM, (b) the 4-state constrained HMM, and (c) the constrained 4-state HHMM. Pink vertical lines indicate when 30 of the Oryx were found dead (note: the remaining 2 dead individuals died shortly after the rainy season, and their death dates were not included).

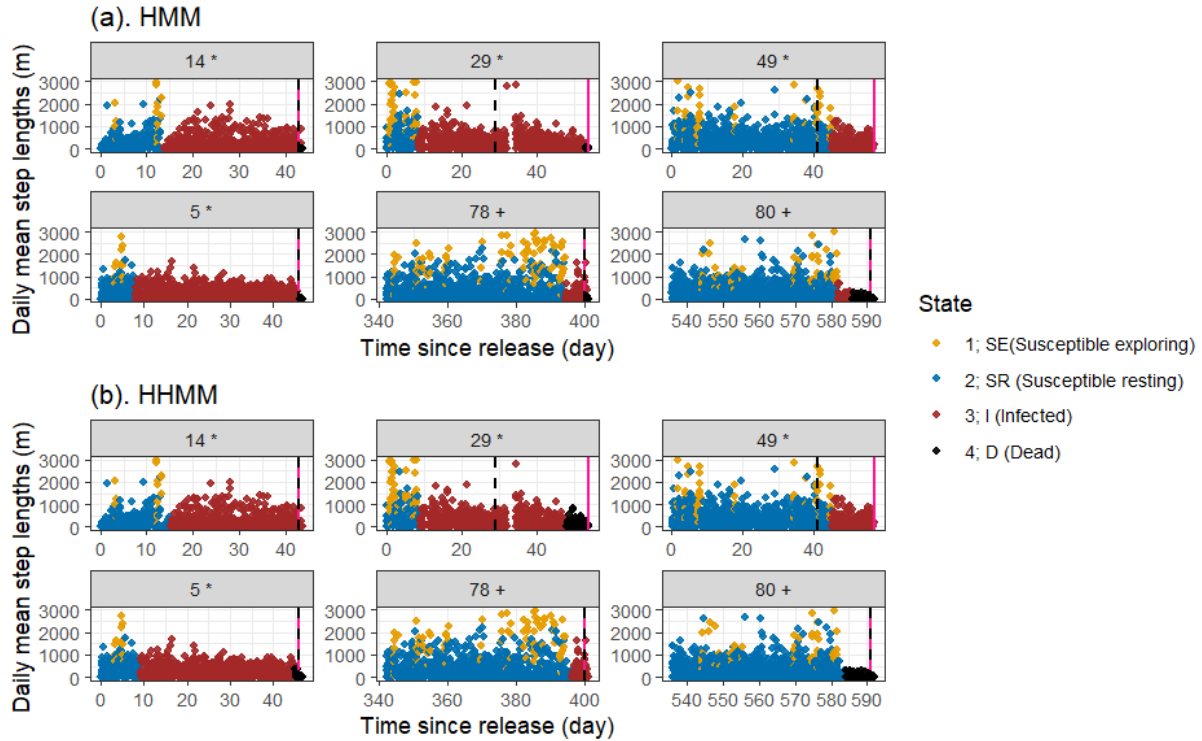

**Figure 6.** Time series of daily mean observed step lengths of oryx that were tested for disease ($n = 6$) with color indicating inferred states from **(a)** the constrained 4-state HMM and **(b)** the constrained 4-state HHMM. Black vertical lines indicate when individuals were tested for disease by veterinarians, and pink vertical lines represent when the animals were found dead (not necessarily the day they died). **(a).** The shape of the symbol at the end of the ID indicates the body condition observed in the field at the last observation (* indicates under body condition index, and + represents optimal body condition index). Individuals 29 and 49 were recaptured and tested for disease status while alive, whereas the other individuals (5, 14, 78, 80) were found dead and tested for disease status during autopsies. Four individuals (5, 14, 29, 49) were released during the 2018 rainy season, and individuals 78 and 80 were released in 2017.

**Data availability statement**: Code is currently available in the GitHub repository at https://github.com/kimx3725/Disease_Oryx.

**Acknowledgments:** The reintroduction of oryx in Chad is a joint initiative of the Government of Chad and the Environment Agency - Abu Dhabi (EAD). Under the overall leadership and management of EAD, on-the-ground implementation is carried out by Sahara Conservation. The Smithsonian's National Zoo and Conservation Biology Institute extends its gratitude to the Government of Chad, EAD, and Sahara Conservation for the opportunity to assist in oryx reintroduction activities. We thank the Environment Agency Abu Dhabi for originating and managing the world herd of scimitar-horned oryx. We are grateful to all current and former Sahara Conservation personnel dedicated to restoring oryx to Chad, including John Newby, Tim Woodfine, Violeta Barrios, Marc Dethier, Habib Ali Hamit, Kher Issakha, Oumar Annadif, Mahamat Ali, Loutfallah Ali, François Madjitigal, Ahmat Anour, Evariste Djibkebeng, Yacoub Mahamat, Hissein Gadeye, Honoré Todjibaye Midjigue, Dieudonné Kephas Doldiguim, Oualbadet Magomna, Daniel Nahodjingar, Ouchar Ahmat, Caleb Ngaba Waye Taroum, Firmin Dingamtebeye, Hiti Ngarya Nouba, and Krazidi Abeye. We thank the Ministère de l'Environnement de la Pêche et du Développement Durable and the Institut de Recherche en Elevage pour le Développement for providing invaluable support to the oryx reintroduction project in Chad. We are grateful to the efforts of the Coordonateur de la Reserve, Mahamat Hassan Hatcha, the Chef de Secteur Nord, the Chef de Secteur Sud, and the entire ranger force of the Direction de la Conservation de la Faune et des Aires Protégées. We also thank Tim Wacher, Justin Chuven, Jon Lllona Minguez, Elena Pesci, Adam Eyres, Julie Swenson, Ben Jernigan, Dan Beetem, Gavin Livingston, RoxAnna Breitigan, and many others who have contributed their valuable time and expertise to the oryx reintroduction project. K.M. was supported by EAD and Sahara Conservation. D.K. was supported by an INTERN supplement on Grant No. DEB-1654609, the Smithsonian Institution Fellowship Program (SIFP), and a Doctoral Dissertation Fellowship from the University of Minnesota. J.F. received partial support from the Minnesota Agricultural Experimental Station. K.M. was supported by a grant from Sahara Conservation and the Environment Agency - Abu Dhabi.

# Supporting Information

Additional supporting information can be found online in the Supporting Information section at the end of this article.

**Appendix S1**

**Detecting disease progression from animal movement using hidden Markov models**

Dongmin Kim, Théo Michelot, Katherine Mertes, Jared Stabach, John Fieberg

## Section S1. Oryx Data Summary

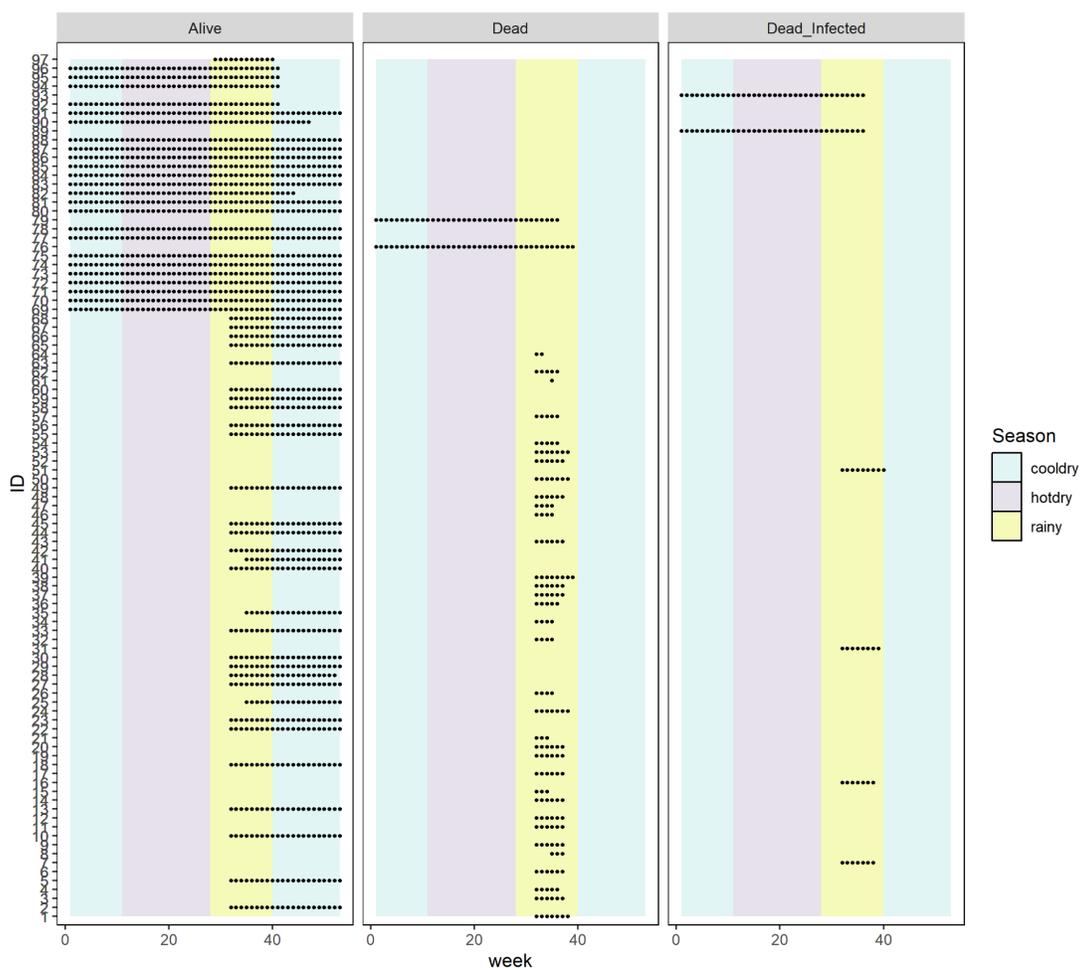

**Figure S1. Summary of Oryx movement data availability.** The study population of 84 tracked oryx included 46 susceptible alive, 32 susceptible (dead), and 6 known infected (dead infected) individuals. Black dots indicate the availability of movement data for a given oryx during a given

week of 2018. Colored boxes indicate the three seasons characteristic of central Chad. Our study period included only the 2018 rainy season, when the majority of released oryx (Release Group 4) died within two months of release.

**Table S1. Summary of study population.** Here we present disease classification (Alive, Dead presumed infected, and Dead confirmed infected), release group, date of arrival to the RFOROA, date each individual was fit with a tracking device, date of release into the reserve, and date of death (based on alerts from tracking devices and field observations) for the 84 oryx in our study population.

| ID | Disease classification | Release group | Release site arrival date | Date collared | Release date | Death date |
|---|---|---|---|---|---|---|
| 1 | Dead | 4 | 02/12/2018 | 07/22/2018 | 08/06/2018 | 9/20/2018 |
| 2 | Alive | 4 | 02/19/2018 | 07/22/2018 | 08/06/2018 | 6/10/2019 |
| 3 | Alive | 4 | 02/12/2018 | 07/21/2018 | 08/06/2018 | NA |
| 4 | Dead | 4 | 02/16/2018 | 07/21/2018 | 08/06/2018 | 9/15/2018 |
| 5 | Dead_Infected | 4 | 02/12/2018 | 07/22/2018 | 08/06/2018 | 9/24/2018 |
| 6 | Dead | 4 | 02/19/2018 | 07/21/2018 | 08/27/2018 | 9/12/2018 |
| 7 | Dead | 4 | 02/12/2018 | 07/21/2018 | 08/06/2018 | 9/16/2018 |
| 8 | Alive | 4 | 02/16/2018 | 07/22/2018 | 08/06/2018 | 2/25/2020 |
| 9 | Dead | 4 | 02/19/2018 | 7/22/2018 | 8/6/2018 | 9/15/2018 |
| 10 | Dead | 4 | 02/16/2018 | 07/22/2018 | 08/06/2018 | 9/13/2018 |
| 11 | Alive | 4 | 02/12/2018 | 07/22/2018 | 08/06/2018 | NA |
| 12 | Dead | 4 | 02/16/2018 | 07/21/2018 | 08/06/2018 | 9/16/2018 |
| 13 | Dead | 4 | 02/12/2018 | 07/21/2018 | 08/06/2018 | 8/26/2018 |
| 14 | Dead_Infected | 4 | 02/16/2018 | 07/22/2018 | 08/06/2018 | 9/18/2018 |
| 15 | Dead | 4 | 02/19/2018 | 07/22/2018 | 08/06/2018 | 9/25/2018 |
| 16 | Alive | 4 | 02/16/2018 | 07/22/2018 | 08/06/2018 | NA |
| 17 | Dead | 4 | 02/12/2018 | 07/21/2018 | 08/06/2018 | 9/13/2018 |
| 18 | Dead | 4 | 02/19/2018 | 07/21/2018 | 08/06/2018 | 9/15/2018 |

| 19 | Dead | 4 | 02/16/2018 | 07/22/2018 | 08/06/2018 | 8/22/2018 |
|----|------|---|------------|------------|------------|-----------|
| 20 | Alive | 4 | 02/16/2018 | 07/21/2018 | 08/06/2018 | NA |
| 21 | Alive | 4 | 02/12/2018 | 07/21/2018 | 08/06/2018 | NA |
| 22 | Dead | 4 | 02/19/2018 | 07/21/2018 | 08/06/2018 | 9/21/2018 |
| 23 | Alive | 4 | 02/19/2018 | 07/21/2018 | 09/01/2018 | NA |
| 24 | Dead | 4 | 02/19/2018 | 07/21/2018 | 08/06/2018 | 9/2/2018 |
| 25 | Alive | 4 | 02/12/2018 | 07/21/2018 | 08/06/2018 | NA |
| 26 | Alive | 4 | 02/16/2018 | 07/21/2018 | 08/06/2018 | NA |
| 27 | Alive | 4 | 02/12/2018 | 07/22/2018 | 08/06/2018 | NA |
| 28 | Alive | 4 | 02/16/2018 | 07/21/2018 | 08/06/2018 | NA |
| 29 | Dead_Infected | 4 | 02/19/2018 | 07/21/2018 | 08/06/2018 | 9/29/2018 |
| 30 | Dead | 4 | 02/19/2018 | 07/22/2018 | 08/06/2018 | 8/30/2018 |
| 31 | Alive | 4 | 02/19/2018 | 07/21/2018 | 08/06/2018 | NA |
| 32 | Dead | 4 | 02/16/2018 | 07/21/2018 | 08/06/2018 | 8/27/2018 |
| 33 | Alive | 4 | 02/19/2018 | 07/21/2018 | 09/02/2018 | NA |
| 34 | Dead | 4 | 02/12/2018 | 07/22/2018 | 08/06/2018 | 9/3/2018 |
| 35 | Dead | 4 | 02/12/2018 | 07/22/2018 | 08/06/2018 | 9/12/2018 |
| 36 | Dead | 4 | 02/16/2018 | 07/21/2018 | 08/06/2018 | 9/11/2018 |
| 37 | Dead | 4 | 02/19/2018 | 07/22/2018 | 08/06/2018 | 9/24/2018 |
| 38 | Alive | 4 | 02/16/2018 | 07/21/2018 | 08/06/2018 | NA |
| 39 | Alive | 4 | 02/16/2018 | 07/22/2018 | 09/01/2018 | NA |
| 40 | Alive | 4 | 02/16/2018 | 07/21/2018 | 08/06/2018 | 6/24/2020 |
| 41 | Dead | 4 | 02/19/2018 | 07/22/2018 | 08/06/2018 | 9/11/2018 |
| 42 | Alive | 4 | 02/16/2018 | 07/22/2018 | 08/06/2018 | NA |
| 43 | Alive | 4 | 02/12/2018 | 07/22/2018 | 08/06/2018 | NA |
| 44 | Dead | 4 | 02/16/2018 | 07/22/2018 | 08/06/2018 | 8/29/2018 |
| 45 | Dead | 4 | 02/12/2018 | 07/21/2018 | 08/06/2018 | 8/30/2018 |

| 46 | Dead | 4 | 02/16/2018 | 07/21/2018 | 08/06/2018 | 9/14/2018 |
|----|------|---|------------|------------|------------|-----------|
| 47 | Alive | 4 | 02/12/2018 | 07/22/2018 | 08/06/2018 | NA |
| 48 | Dead | 4 | 02/12/2018 | 07/21/2018 | 08/06/2018 | 9/21/2018 |
| 49 | Dead_Infected | 4 | 02/12/2018 | 07/22/2018 | 08/06/2018 | 10/2/2018 |
| 50 | Dead | 4 | 02/16/2018 | 07/22/2018 | 08/06/2018 | 9/14/2018 |
| 51 | Dead | 4 | 02/12/2018 | 07/21/2018 | 08/06/2018 | 9/19/2018 |
| 52 | Dead | 4 | 02/19/2018 | 07/22/2018 | 08/06/2018 | 9/7/2018 |
| 53 | Alive | 4 | 02/19/2018 | 07/22/2018 | 08/06/2018 | NA |
| 54 | Alive | 4 | 02/16/2018 | 07/22/2018 | 08/06/2018 | NA |
| 55 | Dead | 4 | 02/19/2018 | 07/22/2018 | 08/06/2018 | 9/8/2018 |
| 56 | Alive | 4 | 02/16/2018 | 07/22/2018 | 08/06/2018 | NA |
| 57 | Alive | 4 | 02/16/2018 | 07/22/2018 | 08/06/2018 | NA |
| 58 | Alive | 4 | 02/19/2018 | 07/21/2018 | 08/06/2018 | NA |
| 59 | Dead | 4 | 02/12/2018 | 07/21/2018 | 08/28/2018 | 9/1/2018 |
| 60 | Dead | 4 | 02/12/2018 | 07/22/2018 | 08/06/2018 | 9/4/2018 |
| 61 | Alive | 4 | 02/12/2018 | 07/22/2018 | 08/06/2018 | NA |
| 62 | Dead | 4 | 02/16/2018 | 07/21/2018 | 08/06/2018 | 8/21/2018 |
| 63 | Alive | 4 | 02/16/2018 | 07/22/2018 | 08/06/2018 | NA |
| 64 | Alive | 4 | 02/19/2018 | 07/21/2018 | 08/06/2018 | NA |
| 65 | Alive | 4 | 02/12/2018 | 07/22/2018 | 08/06/2018 | NA |
| 66 | Alive | 4 | 02/19/2018 | 7/22/2018 | 8/6/2018 | NA |
| 67 | Alive | 3 | 01/19/2017 | 7/31/2017 | 9/25/2017 | 3/2/2021 |
| 68 | Alive | 3 | 01/19/2017 | 7/31/2017 | 8/3/2017 | NA |
| 69 | Alive | 3 | 01/19/2017 | 7/31/2017 | 8/3/2017 | NA |
| 70 | Alive | 3 | 01/19/2017 | 7/31/2017 | 8/3/2017 | NA |
| 71 | Alive | 3 | 01/19/2017 | 7/31/2017 | 8/3/2017 | NA |
| 72 | Alive | 3 | 01/19/2017 | 7/31/2017 | 8/3/2017 | NA |

| 73 | Alive | 3 | 11/16/2016 | 10/11/2016 | 8/3/2017 | 10/31/2018 |
|---|---|---|---|---|---|---|
| 74 | Alive | 3 | 01/19/2017 | 7/31/2017 | 8/3/2017 | NA |
| 75 | Alive | 3 | 01/19/2017 | 7/31/2017 | 8/3/2017 | NA |
| 76 | Alive | 3 | 01/19/2017 | 7/31/2017 | 8/3/2017 | NA |
| 77 | Alive | 3 | 11/16/2016 | 7/31/2017 | 8/3/2017 | NA |
| 78 | Dead_Infected | 3 | 01/19/2017 | 7/31/2017 | 8/3/2017 | 9/7/2018 |
| 79 | Alive | 2 | 11/16/2016 | 10/11/2016 | 1/21/2017 | NA |
| 80 | Dead_Infected | 2 | 11/16/2016 | 10/11/2016 | 1/21/2017 | 9/4/2018 |
| 81 | Alive | 2 | 11/16/2016 | 10/11/2016 | 1/21/2017 | NA |
| 82 | Alive | 2 | 11/16/2016 | 10/11/2016 | 1/21/2017 | NA |
| 83 | Alive | 2 | 11/16/2016 | 10/11/2016 | 1/21/2017 | NA |
| 84 | Alive | 2 | 11/16/2016 | 10/11/2016 | 1/21/2017 | NA |

**Table S2. Summary of the six infected dead individuals.** Most of the individuals were co-infected with several pathogen and parasite diseases such as Rift Valley Fever (RVF), Peste des Petits Ruminants (PPR), Hemorrhagic septicemia, Pasteurellosis, and Babesiosis. Some were found dead and underwent an autopsy, while two individuals (29 and 49) were recaptured for further testing while alive.

| ID | Testing methods | Infections |
|----|-----------------|------------|
| 5 | Autopsy | RVF, PPR, Hemorrhagic septicemia |
| 14 | Autopsy | Hemorrhagic septicemia |
| 29 | Recapture and vet tests (PCR) | RVF, Babesiosis |
| 49 | Recapture and vet tests (PCR) | RVF, Babesiosis |
| 78 | Autopsy | PPR, Hemorrhagic septicemia |
| 80 | Autopsy | Pasteurellosis, RVF, PPR, Babesiosis, Hemorrhagic septicemia |

**Section S2. Simulation Data Summary**

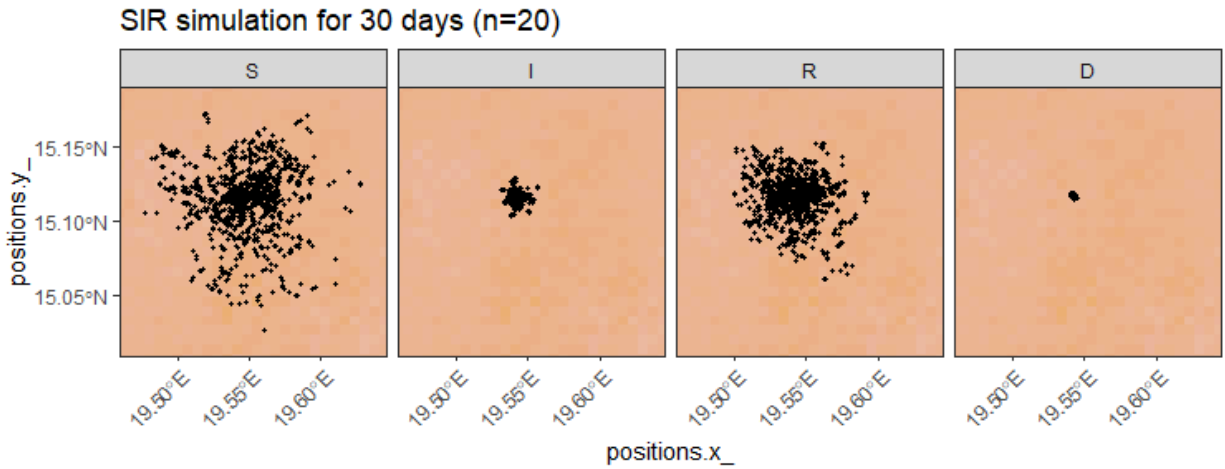

**Figure S2.** Locations of simulated movement tracks of 20 individuals for 30 days. Each dot indicates the hourly location of the simulated individuals. The simulated individuals move differently based on their daily infection-related states. Susceptible (S) and recovered (R) individuals move similarly. In contrast, infected (I) individuals move less and dead (D) do not move.

**Section S3. Time series of mean daily step length.**

We plotted mean daily step length versus time since release for all oryx in the study population (up to 90 days). We resampled oryx movement data to hourly locations to create a series of steps with a constant sampling interval and calculated the mean step length per day.

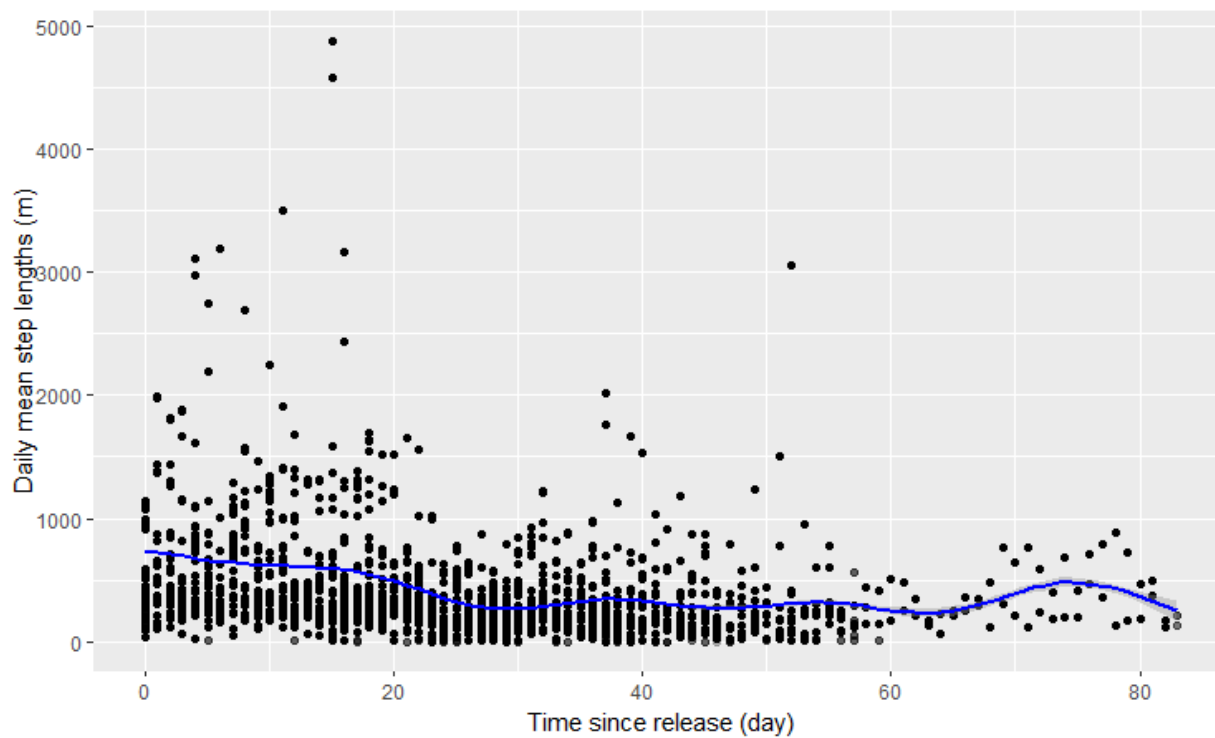

**Figure S3.** Time series of daily mean step lengths since the time release from 84 tagged individuals. The blue line indicates the smooth line of the total daily mean step lengths across the tagged individuals.

**Appendix S2**

**Detecting disease progression from animal movement using hidden Markov models**

Dongmin Kim, Théo Michelot, Katherine Mertes,  Jared Stabach, John Fieberg

**Section S1. Choosing initial parameter values for the estimation**

To avoid convergence issues in the likelihood optimization, we fit 15 models per state model (i.e., 3, 4, and 5 state HMMs) with different sets of starting values and select the best model fit among those based on log-likelihood values.

The best model per state is listed below with their maximum log-likelihood value among 15 model runs:

<u>3-state HMM</u>:

```
Maximum log-likelihood values:
-833747.1 -822614.4 -822614.4 -822614.4 -833747.1 -822614.4 -822614.4
-822614.4 -822614.4 -822614.4 -822614.4 -822614.4 -822614.4 -822614.4
-822614.4

Best model:

Value of the maximum log-likelihood: -822614.4

step parameters:
----------------
      state 1  state 2  state 3
mean 4.226161 317.6875 1809.043
sd   3.159653 364.1812 1064.134

angle parameters:
-----------------
               state 1    state 2     state 3
mean          3.1383642 0.07831160 -0.0368014
concentration 0.4200796 0.08711126  1.5305686
```

<u>4-state SI HMM</u>:

```
Maximum log-likelihood values:
-819189.6 -819189.6 -819189.6 -819189.6 -819189.6 -819189.6 -819189.7
-819189.6 -819189.7 -819189.6 -819189.6 -819189.6 -819189.6 -819189.6
-819189.6
```

```
Best model:
Value of the maximum log-likelihood: -819189.6

step parameters:
----------------
       state 1  state 2  state 3  state 4
mean 2.680949 21.77835 382.0162 1945.086
sd   1.513453 19.21009 369.3465 1037.136

angle parameters:
----------------
               state 1     state 2    state 3     state 4
mean         3.1337626 -3.1320171  0.0389272 -0.03153871
concentration 0.3302889  0.3628805  0.1696356  1.55942826
```

## 4-state SIR HMM:

```
Maximum log-likelihood values:
-844156.4, -844156.5, -844156.4, -844156.4, -844156.4, -844156.4, -844156.4,
-844156.4, -844156.4, -844156.4, -844156.4, -844156.4, -844156.4, -844156.4,
-844156.4

Best model:
Value of the maximum log-likelihood: -844156.4

step parameters:
----------------
          S        I        R        D
mean  761.2258 299.5161 750.1033 14.93762
sd   1103.5737 442.1466 484.1662 19.47883

angle parameters:
----------------
                  S          I            R          D
mean        -0.01995334 0.24192066 -0.007260016 -3.1387158
concentration 0.41713787 0.01655404  0.010008986  0.5741106
```

## 5-state HMM:

```
Maximum log-likelihood values:
-817441.4 -817441.4 -817441.4 -817441.4 -817441.4 -817441.4 -817441.4
-817441.4 -817441.4 -817441.4 -817441.4 -817441.4 -817441.4 -817447.0
-817441.4

Best model:
Value of the maximum log-likelihood: -817441.4

step parameters:
----------------
      state 1  state 2  state 3  state 4   state 5
```

```
mean 2.524012 15.25322 220.6390 751.7033 2466.1359
sd   1.364508 12.17583 221.5885 484.0975  880.9561

angle parameters:
-----------------
              state 1    state 2   state 3     state 4      state 5
mean         3.134879 -3.1259356 0.42186268 -0.008988095 -0.02347248
concentration 0.326152  0.3690583 0.01808188  0.531983987  1.80426823
```

**Section S2. Choosing an appropriate number of states and model diagnostics**

To check whether models with a different number of states change the model predictions, we fitted HMMs with three different sets of states (i.e., 3-state HMM, 4-state HMM, and 5-state HMM). We further checked whether the restrictions on the model transition probability matrix based on contemporaneous *in situ* observations showed more realistic decoded states predicted by the models. Thus, we first fitted an HMM without restrictions on the transition probability matrix for models with 3, 4, and 5 states. We second fitted an HMM with restrictions on the transition probability matrix for each state model. We then plotted a time series of step lengths colored by the decoded states from the models to check if both models provided similar decoded states (Figures S2 and S5). We observed that the decoded states were different from each other. We found that the 4-state model's decoded states were more reasonable than the 3- and 5-state models (Figures S2 and S5). For example, the constrained 4-state HMM correctly classified 33 of the 38 confirmed dead individuals, whereas the constrained 3 or 5-state HMM correctly classified fewer individuals than the 4-state HMM. Thus, we selected the 4-state model as our final model for the study.

We also plotted pseudo-residuals based on the estimated step lengths (top plots) and turning angles (bottom plots) from the 3-state HMM (Figure S4), the 4-state HMM (Figure S5) and 5-state HMM (Figure S6), and 4-state HHMM (Figure S7): (1). Time series plot of pseudo-residuals (left panel), (2). Q-Q plot of pseudo-residuals (middle panel), and (3) autocorrelation function (ACF) plot (right panel).

All models show similar trends from the time series plots, with residuals fluctuating randomly over time. We see some deviations in the lower tail of the step length distribution. The model predicts more very short step lengths than what is observed in the actual data. This may indicate the possibility of measurement error. Compared to the (H)HMMs, the 4-state constrained HMM appears to fit slightly better to the normal distribution, although some extreme values still deviate (Figure S5).

## Section S3. 3-state HMMs

## 3.1. Unconstrained 3-state model

```
Value of the maximum log-likelihood: -837084.8

step parameters:
----------------
            S         I         D
mean 1450.084 229.9058 15.61235
sd   1072.560 331.9978 20.55279

angle parameters:
-----------------
                     S          I          D
mean        -0.04416724 2.98623791 -3.1334234
concentration 1.34392673 0.06907465  0.5466987

Regression coeffs for the transition probabilities:
---------------------------------------------------
                      1 -> 2        1 -> 3        2 -> 1      2 -> 3 3 -> 1
3 -> 2
(Intercept)        -0.9802697513 -6.6575337124 -2.4334206409 -1.5677767 -1e+06
-1e+06
shrub              -0.0152683936  0.0496176957 -0.0627233723 -0.4948396 -1e+06
-1e+06
time_since_release -0.0004278216  0.0008501638 -0.0003082179 -1.5771796 -1e+06
-1e+06

Transition probability matrix (based on mean covariate values):
---------------------------------------------------------------
            S         I            D
S 0.75985182 0.2385520  1.596233e-03
I 0.05061246 0.9493875  1.278511e-113
D 0.00000000 0.0000000  1.000000e+00

Initial distribution:
---------------------
            S            I            D
3.864444e-11 1.000000e+00 1.188900e-16
```

## 3.2. Constrained 3-state model

```
Value of the maximum log-likelihood: -845574.2

step parameters:
----------------
            S         I         D
mean   745.2912 294.7914   9.548312
```

```
sd   1087.3869 439.8388 11.125256

angle parameters:
-----------------
                     S          I        D
mean          -0.01798164 0.711546828 2.913282
concentration  0.40192371 0.004641055 0.702596

Regression coeffs for the transition probabilities:
---------------------------------------------------
                      1 -> 2 1 -> 3 2 -> 1    2 -> 3 3 -> 1 3 -> 2
(Intercept)       -6.8968735682 -1e+06 -1e+06 -4.6815411 -1e+06 -1e+06
shrub              0.0533211177 -1e+06 -1e+06  0.2317948 -1e+06 -1e+06
time_since_release 0.0005727067 -1e+06 -1e+06 -9.4985281 -1e+06 -1e+06

Transition probability matrix (based on mean covariate values):
--------------------------------------------------------------
          S          I D
S 0.9983791 0.001620881 0
I 0.0000000 1.000000000 0
D 0.0000000 0.000000000 1

Initial distribution:
---------------------
          S          I          D
6.761259e-01 3.238448e-01 2.925778e-05
```

### 3.3 Constrained 3-state model with known death <span style="color:red">(no convergence)</span>

```
Value of the maximum log-likelihood: -1.797693e+308

step parameters:
----------------
        S   I D
mean 1800 315 4
sd   1050 365 4

angle parameters:
-----------------
              S    I        D
mean       -0.03 0.07 3.141593
concentration  1.50 0.08 0.400000

Regression coeffs for the transition probabilities:
---------------------------------------------------
                   1 -> 2 1 -> 3 2 -> 1 2 -> 3 3 -> 1 3 -> 2
(Intercept)         -1.5 -1e+06 -1e+06   -1.5 -1e+06 -1e+06
shrub                0.0 -1e+06 -1e+06    0.0 -1e+06 -1e+06
time_since_release   0.0 -1e+06 -1e+06    0.0 -1e+06 -1e+06
```

```
Transition probability matrix (based on mean covariate values):
--------------------------------------------------------------
          S          I          D
S 0.8175745 0.1824255 0.0000000
I 0.0000000 0.8175745 0.1824255
D 0.0000000 0.0000000 1.0000000

Initial distribution:
---------------------
        S          I          D
0.3333333 0.3333333 0.3333333
```

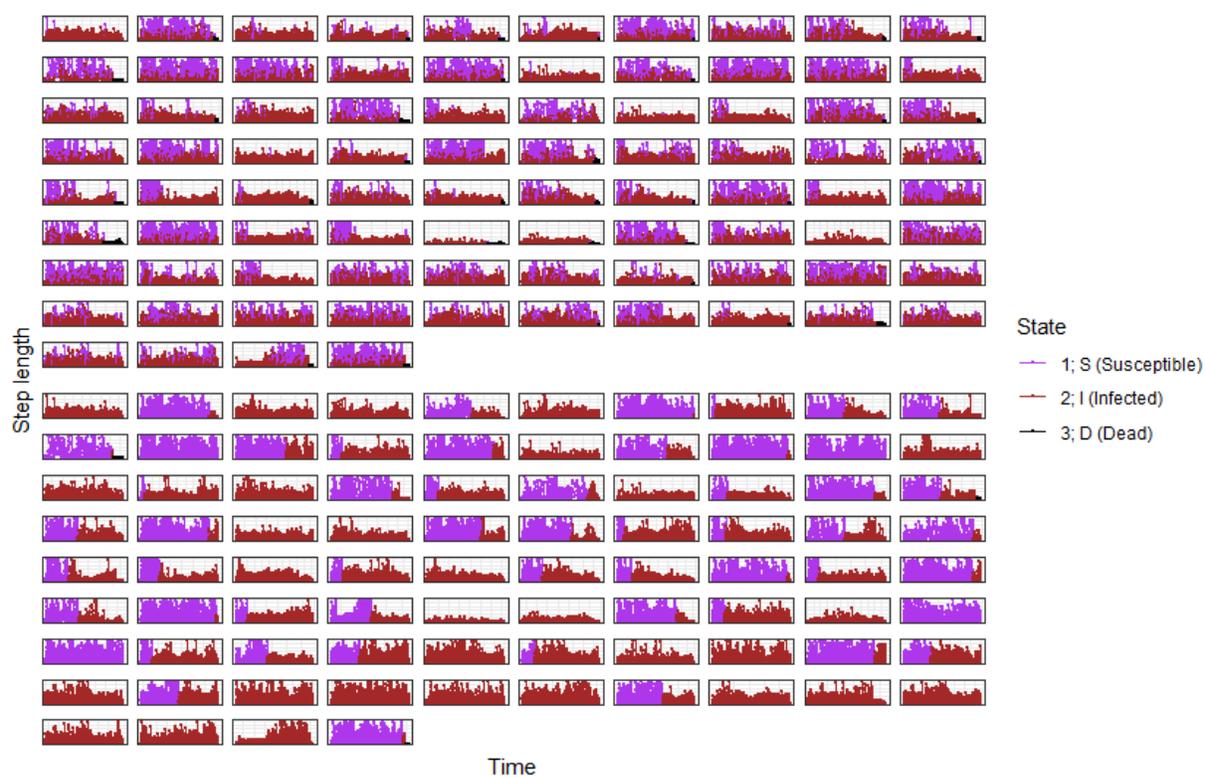

**Figure S2.** A time series of step lengths colored by the decoded states from the unconstrained 3-state HMM (a top panel) vs the constrained 3-state HMM (a bottom panel)

## Section S4. 4-state (H)HMMs

### 4.1. Unconstrained 4-state model (SIS)

```
Value of the maximum log-likelihood: -825583.4

step parameters:
----------------
          SE      SR       I        D
mean 1802.752 317.9315 4.160769 8.466996
sd   1077.595 363.8234 3.086291 8.574073

angle parameters:
-----------------
                        SE         SR         I          D
mean           -0.03640715 0.08437032 3.1408276 3.0658192
concentration   1.52331974 0.08402688 0.4032178 0.7689519

Regression coeffs for the transition probabilities:
---------------------------------------------------
                        1 -> 2        1 -> 3     1 -> 4        2 -> 1        2 -> 3
(Intercept)        -0.9506075888 -3.7405203961 -2.391813 -2.9138415733 -2.2063569020
shrub              -0.0264951454  0.0744572504 -4.895373 -0.0255400360  0.0248192845
time_since_release -0.0003633459 -0.0004985664 -6.008808 -0.0001662611 -0.0006136597
                        2 -> 4        3 -> 1        3 -> 2         3 -> 4  4 -> 1  4 -> 2
(Intercept)          -1.774485 -2.7063824545 -0.030504019 -1.8038954963 -1e+06 -1e+06
shrub                -1.986169 -0.0733494976 -0.036589146 -0.6715370716 -1e+06 -1e+06
time_since_release   -2.366991 -0.0009999114 -0.000153738  0.0005477064 -1e+06 -1e+06
                        4 -> 3
(Intercept)          -1e+06
shrub                -1e+06
time_since_release   -1e+06

Transition probability matrix (based on mean covariate values):
---------------------------------------------------------------
         SE         SR         I           D
SE 0.74689275 0.2252484 0.02785886  0.000000e+00
SR 0.03784118 0.8597858 0.10237304  8.535314e-174
I  0.01907092 0.4130374 0.56705285  8.388240e-04
D  0.00000000 0.0000000 0.00000000  1.000000e+00

Initial distribution:
---------------------
          SE         SR         I          D
0.0114903337 0.0201674303 0.9673900647 0.0009521713
```

### 4.2. Constrained 4-state model (SI)

```
Value of the maximum log-likelihood: -835806.7

step parameters:
----------------
          SE      SR       I        D
mean 1629.186 275.4228 194.0144 14.05722
sd   1094.530 391.6293 285.0358 18.09655
```

```
angle parameters:
-----------------
                       SE          SR           I           D
mean          -0.03771452 3.02589153 2.79016429 -3.1302463
concentration  1.55017641 0.03465011 0.04698539  0.5922272

Regression coeffs for the transition probabilities:
---------------------------------------------------
                          1 -> 2 1 -> 3 1 -> 4      2 -> 1      2 -> 3   2 -> 4
(Intercept)       -0.9519603969  -1e+06  -1e+06 -2.7072246135 -6.762835566 -9.042585
shrub             -0.0122131328  -1e+06  -1e+06 -0.0214502286 -0.061725601  0.000000
time_since_release -0.0002766136  -1e+06  -1e+06  0.0006709467  0.001667105  0.000000
                          3 -> 1 3 -> 2    3 -> 4 4 -> 1 4 -> 2 4 -> 3
(Intercept)       -1e+06  -1e+06 -6.864947  -1e+06  -1e+06  -1e+06
shrub             -1e+06  -1e+06  0.000000  -1e+06  -1e+06  -1e+06
time_since_release -1e+06  -1e+06  0.000000  -1e+06  -1e+06  -1e+06

Transition probability matrix (based on mean covariate values):
---------------------------------------------------------------
           SE         SR          I            D
SE 0.74724223 0.2527578 0.0000000000 0.0000000000
SR 0.05990208 0.9390728 0.0009140226 0.0001110592
I  0.00000000 0.0000000 0.9989573502 0.0010426498
D  0.00000000 0.0000000 0.0000000000 1.0000000000

Initial distribution:
---------------------
          SE          SR          I            D
2.558638e-06 9.472805e-01 5.271689e-02 4.163098e-09
```

## 4.3. Constrained 4-state model (SIR)

```
Value of the maximum log-likelihood: -844156.4

step parameters:
----------------
           S        I          R        D
mean  761.2442 299.5190    14.88417 14.93824
sd   1103.5958 442.1514 15953.56265 19.47979

angle parameters:
-----------------
                       S          I          R           D
mean          -0.01994077 0.24167270 3.1148948 -3.1387241
concentration  0.41714353 0.01655916 0.1037438  0.5741298

Regression coeffs for the transition probabilities:
---------------------------------------------------
                          1 -> 2 1 -> 3 1 -> 4 2 -> 1   2 -> 3    2 -> 4 3 -> 1 3
-> 2 3 -> 4 4 -> 1 4 -> 2 4 -> 3
(Intercept)       -6.9964291956 -1e+06 -1e+06 -1e+06  0.700000 -7.572677 -1e+06
-1e+06 -1e+06 -1e+06 -1e+06 -1e+06
shrub              0.0876829872 -1e+06 -1e+06 -1e+06 -6.878958  0.000000 -1e+06
-1e+06 -1e+06 -1e+06 -1e+06 -1e+06
```

```
time_since_release  0.0002299698 -1e+06 -1e+06 -1e+06 -4.678367  0.000000 -1e+06
-1e+06 -1e+06 -1e+06 -1e+06 -1e+06

Transition probability matrix (based on mean covariate values):
--------------------------------------------------------------
          S            I R           D
S 0.9982255 0.001774543 0 0.0000000000
I 0.0000000 0.999485951 0 0.0005140494
R 0.0000000 0.000000000 1 0.0000000000
D 0.0000000 0.000000000 0 1.0000000000

Initial distribution:
---------------------
            S            I            R            D
0.6829887161 0.3162162716 0.0003976476 0.0003973648
```

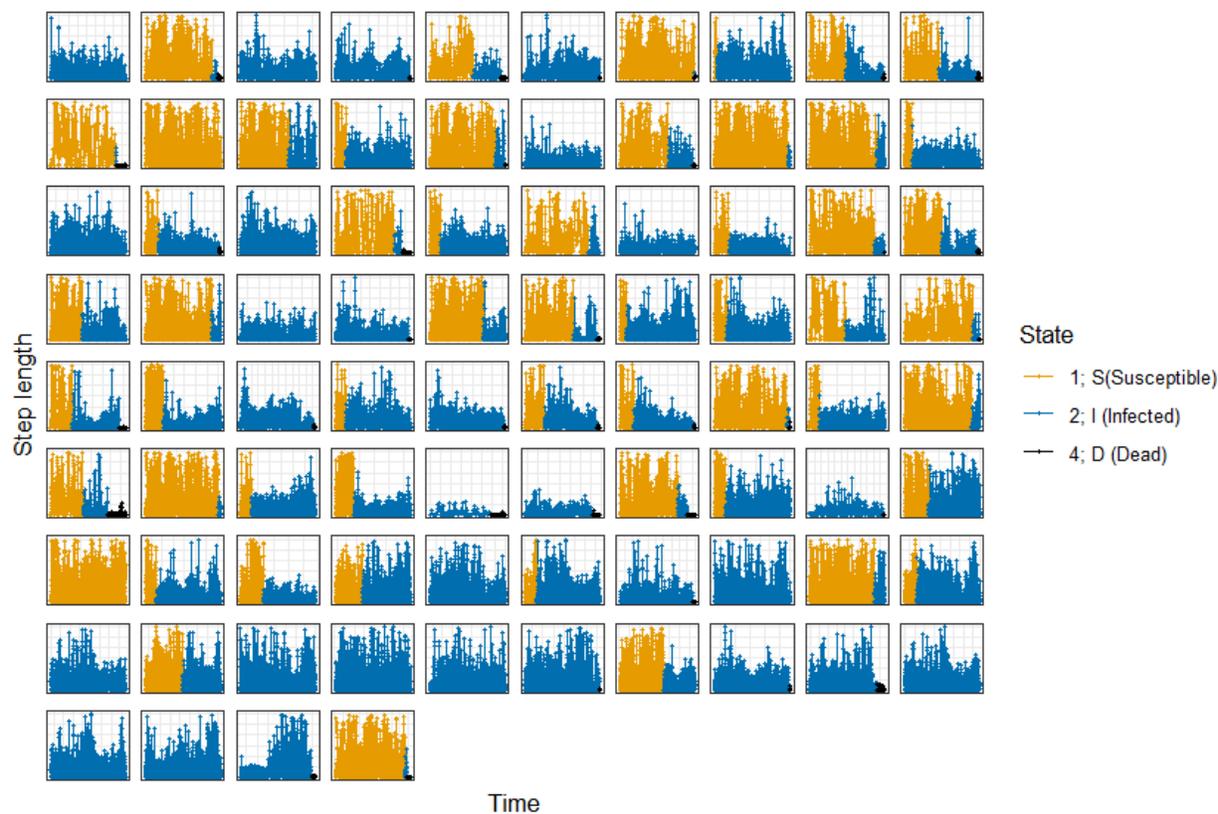

**Figure S5.** A time series of step lengths colored by the decoded states from the constrained 4-state SIR HMM

## 4.4. Constrained 4-state model (SIR) to the simulated data

```
Value of the maximum log-likelihood: -29806.13
```

```
step parameters:
----------------
            S        I        R        D
mean 557.5840  82.6321 503.4475   7.428095
sd   827.2607 185.4812 691.6368  20.833017

angle parameters:
-----------------
                    S          I          R          D
mean          0.01167078 -1.92784911 -0.3532613 0.09077677
concentration 1.68843800  0.02313637  0.1373223 0.55093193

Regression coeffs for the transition probabilities:
---------------------------------------------------
                 1 -> 2 1 -> 3 1 -> 4 2 -> 1    2 -> 3    2 -> 4 3 -> 1 3 -> 2 3
-> 4 4 -> 1 4 -> 2
(Intercept) -4.269345 -1e+06 -1e+06 -1e+06 -6.311916 -4.114572 -1e+06 -1e+06
-1e+06 -1e+06 -1e+06
                 4 -> 3
(Intercept) -1e+06

Transition probability matrix:
------------------------------
          S          I          R          D
S 0.9862021 0.01379789 0.00000000 0.00000000
I 0.0000000 0.98217598 0.00178221 0.01604181
R 0.0000000 0.00000000 1.00000000 0.00000000
D 0.0000000 0.00000000 0.00000000 1.00000000

Initial distribution:
---------------------
            S            I            R            D
9.999989e-01 3.792915e-07 3.792911e-07 3.792908e-07
```

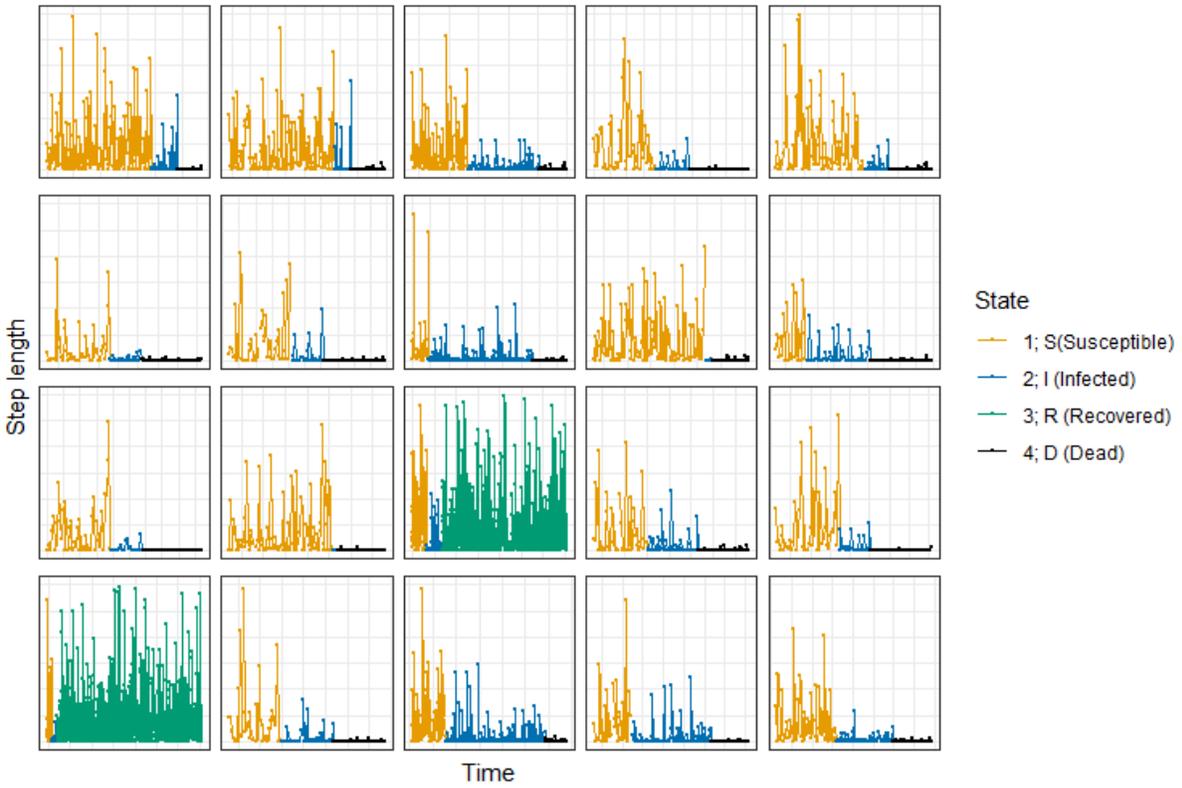

**Figure S5.** A time series of step lengths colored by the decoded states from the constrained 4-state SIR HMM fitted to the simulated data.

**Table S1.** 38 confirmed dead individuals with their disease classifications and the constrained 4-state HMM prediction. The constrained 4-state HMM correctly predicted 33 dead individuals out of the 38 confirmed dead individuals.

| ID | Disease classification | Model prediction |
|----|------------------------|------------------|
| 1 | Dead | Infected |
| 4 | Dead | Dead |
| 5 | Dead | Dead |
| 6 | Dead | Dead |
| 7 | Dead | Dead |
| 9 | Dead | Dead |
| 10 | Dead | Dead |
| 12 | Dead | Dead |
| 13 | Dead | Dead |
| 14 | Dead | Dead |
| 15 | Dead | Dead |
| 17 | Dead | Dead |
| 18 | Dead | Dead |
| 19 | Dead | Dead |
| 22 | Dead | Dead |
| 24 | Dead | Dead |
| 29 | Dead | Dead |
| 30 | Dead | Dead |
| 32 | Dead | Susceptible |
| 34 | Dead | Infected |
| 35 | Dead | Dead |
| 36 | Dead | Dead |
| 37 | Dead | Infected |
| 41 | Dead | Dead |
| 44 | Dead | Infected |
| 45 | Dead | Dead |
| 46 | Dead | Dead |
| 48 | Dead | Dead |
| 49 | Dead | Infected |
| 50 | Dead | Dead |
| 51 | Dead | Dead |
| 52 | Dead | Dead |
| 55 | Dead | Dead |
| 59 | Dead | Dead |
| 60 | Dead | Dead |
| 62 | Dead | Dead |
| 78 | Dead | Dead |
| 80 | Dead | Dead |

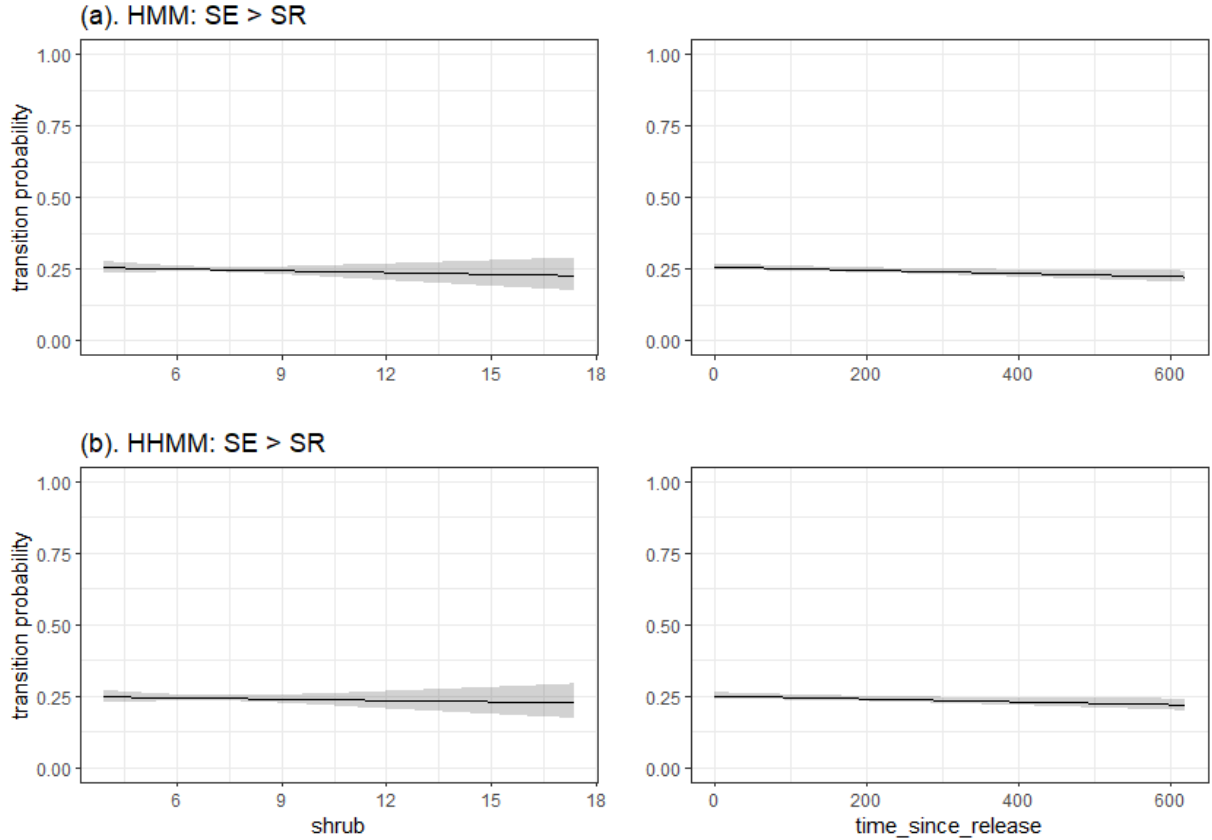

**Figure S3.** Transition probabilities from susceptible exploring (SE) to non-infected resting (SR) as a function of shrub cover and time since release with 95% confidence intervals from **(a)** the 4-state constrained HMM and **(b)** the 4-state constrained HHMM. For both models, there is no clear trend in the transition probabilities from exploring states to resting states as a function of shrub cover when oryx are susceptible (SE to SR). Similarly, there is no trend in the transition probabilities from exploring to resting states as a function of time since release when oryx are susceptible.

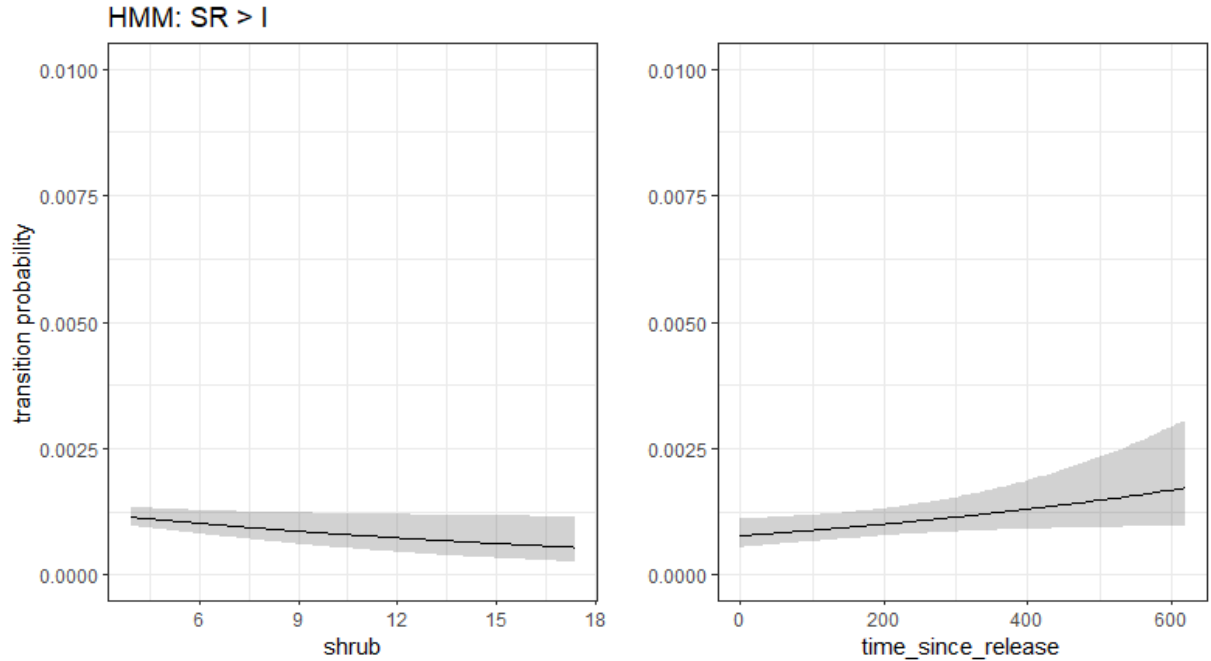

**Figure S4.** Transition probabilities from susceptible resting (SR) to infected (I) as a function of shrub cover and time since release with 95% confidence intervals from the 4-state constrained HMM. There is no trend in the transition probabilities from resting states to infected states as a function of shrub cover and time since release given the wide confidence bands. The y-axis is truncated to help with the visualization of small values.

**Table S2.** 38 confirmed dead individuals with their disease classifications and the constrained 4-state HHMM prediction. The constrained 4-state HHMM correctly predicted 27 dead individuals out of the 38 confirmed dead individuals

| ID | Disease classification | Model prediction |
|---|---|---|
| 1 | Dead | Infected |
| 4 | Dead | Dead |
| 5 | Dead_Infected | Dead |
| 6 | Dead | Dead |
| 7 | Dead | Dead |
| 9 | Dead | Dead |
| 10 | Dead | Dead |
| 12 | Dead | Infected |
| 13 | Dead | Dead |
| 14 | Dead | Infected |
| 15 | Dead | Susceptible |
| 17 | Dead | Dead |
| 18 | Dead | Dead |
| 19 | Dead | Dead |
| 22 | Dead | Infected |
| 24 | Dead | Dead |
| 29 | Dead | Dead |
| 30 | Dead | Dead |
| 32 | Dead | Susceptible |
| 34 | Dead | Dead |
| 35 | Dead | Infected |
| 36 | Dead | Dead |
| 37 | Dead | Dead |
| 41 | Dead | Dead |
| 44 | Dead | Infected |
| 45 | Dead | Susceptible |
| 46 | Dead | Dead |
| 48 | Dead | Dead |
| 49 | Dead | Infected |
| 50 | Dead | Dead |
| 51 | Dead | Dead |
| 52 | Dead | Susceptible |
| 55 | Dead | Dead |
| 59 | Dead | Dead |
| 60 | Dead | Dead |
| 62 | Dead | Infected |
| 78 | Dead | Infected |
| 80 | Dead | Dead |

## 4.3 Constrained 4-state model with known death (no convergence)

```
Value of the maximum log-likelihood: -1.797693e+308

step parameters:
----------------
      SE  SR  I D
mean 1800 380 25 4
sd   1050 370 20 4

angle parameters:
-----------------
               SE    SR    I        D
mean         -0.03  0.03 -3.00 3.141593
concentration 1.50  0.16  0.36 0.300000

Regression coeffs for the transition probabilities:
---------------------------------------------------
                  1 -> 2 1 -> 3 1 -> 4 2 -> 1 2 -> 3 2 -> 4 3 -> 1 3 -> 2 3 -> 4 4
-> 1 4 -> 2 4 -> 3
(Intercept)        -1.5 -1e+06 -1e+06   -1.5   -1.5   -1.5 -1e+06 -1e+06   -1.5
-1e+06 -1e+06 -1e+06
shrub               0.0 -1e+06 -1e+06    0.0    0.0    0.0 -1e+06 -1e+06    0.0
-1e+06 -1e+06 -1e+06
time_since_release  0.0 -1e+06 -1e+06    0.0    0.0    0.0 -1e+06 -1e+06    0.0
-1e+06 -1e+06 -1e+06

Transition probability matrix (based on mean covariate values):
----------------------------------------------------------------
          SE        SR        I         D
SE 0.8175745 0.1824255 0.0000000 0.0000000
SR 0.1336597 0.5990210 0.1336597 0.1336597
I  0.0000000 0.0000000 0.8175745 0.1824255
D  0.0000000 0.0000000 0.0000000 1.0000000

Initial distribution:
---------------------
  SE   SR   I    D
0.25 0.25 0.25 0.25
```

## 4.4 Unconstrained 4-state hierarchical model

```
Value of the maximum log-likelihood: -815708.3

step parameters:
----------------
           e        r        i       id       d       dd
mean 1691.204 297.8768 202.2163 202.2163 60.90892 60.90892
sd   1129.754 419.1183 293.2170 293.2170 97.55779 97.55779

angle parameters:
-----------------
              e        r         i         id        d        dd
mean       -0.03689851 3.135224 -3.13899047 -3.13899047 3.1367153 3.1367153
```

```
concentration
1.57803339 0.000000  0.06929608  0.06929608 0.2634562 0.2634562

------------------------------------------------------------
Regression coeffs for the transition probabilities:
------------------------------------------------------------
------------------------ level1 --------------------------
                    1 -> 3    1 -> 5    3 -> 1    3 -> 5 5 -> 1 5 -> 3
I((level == "1") * 1) -2.175671 -4.901406 -1.956103 -2.862121 -1e+06 -1e+06

------------------------ level2 --------------------------
                                        1 -> 2        2 -> 2
I((level == "2") * 1)                   -0.9868733255  2.7683774306
I((level == "2") * shrub)               -0.0116017860  0.0025923825
I((level == "2") * time_since_release) -0.0003053025 -0.0004443198

                                        3 -> 4 4 -> 4
I((level == "2") * 1)                   -1e+06 -1e+06
I((level == "2") * shrub)               -1e+06 -1e+06
I((level == "2") * time_since_release) -1e+06 -1e+06

                                        5 -> 6 6 -> 6
I((level == "2") * 1)                   -1e+06 -1e+06
I((level == "2") * shrub)               -1e+06 -1e+06
I((level == "2") * time_since_release) -1e+06 -1e+06

------------------------------------------------------------

------------------------------------------------------------
Transition probability matrix (based on mean covariate values):
------------------------------------------------------------
------------------------ level1 --------------------------
            nonInfected  Infected      Death
nonInfected  0.8920861 0.1012803 0.006633662
Infected     0.1179823 0.8343374 0.047680219
Death        0.0000000 0.0000000 1.000000000

------------------------ level2 --------------------------
            e         r
e 0.75382820 0.2461718
r 0.06208636 0.9379136

    i id
i  1  0
id 1  0

    d dd
d  1  0
dd 1  0

------------------------------------------------------------

-----------------------------------------------
Regression coeffs for the initial distribution:
-----------------------------------------------
```

```
-------------------  level1  -------------------
                 state 3   state 5
(Intercept) -14453.31 -18613.09

-------------------  level2  -------------------
                      state 2
I((level == "2i") * 1) 1.630522

                      state 4
I((level == "2i") * 1)        0

                      state 6
I((level == "2i") * 1)        0

-------------------------------------------------
```

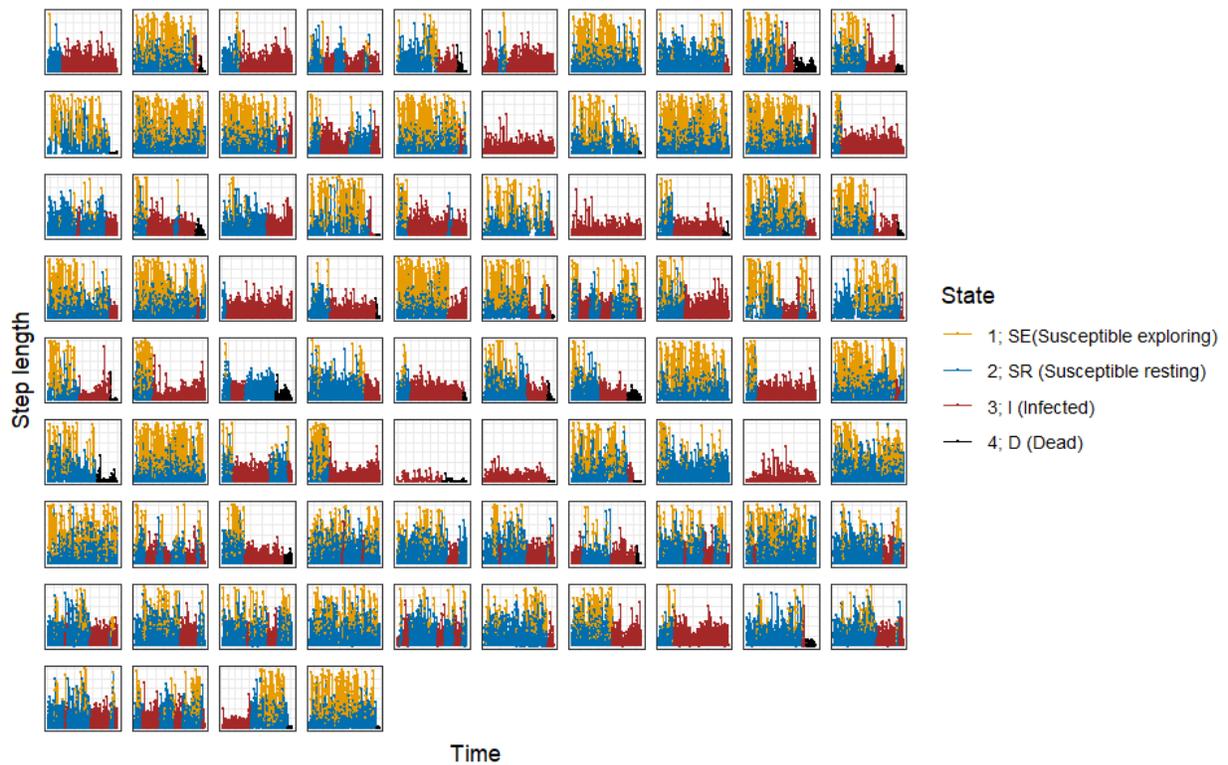

**Figure S6.** A time series of step lengths colored by the decoded states from the unconstrained 4-state HHMM

## 4.5 Constrained 4-state hierarchical model

```
Value of the maximum log-likelihood: -816070.8

step parameters:
----------------
            e       r       i      id       d      dd
mean 1675.549 283.8440 217.3480 217.3480  64.90456  64.90456
sd   1128.406 402.6501 313.2039 313.2039 104.97910 104.97910

angle parameters:
-----------------
                    e            r            i           id          d
dd
mean      -0.03434989 2.381206e+00 -3.00058057 -3.00058057 3.0170807
3.0170807
concentration 1.54178840 9.608558e-10  0.04483596  0.04483596 0.2640674
0.2640674

-------------------------------------------------------------
Regression coeffs for the transition probabilities:
-------------------------------------------------------------
------------------------- level1 -------------------------
                        1 -> 3    1 -> 5 3 -> 1   3 -> 5 5 -> 1 5 -> 3
I((level == "1") * 1) -2.899543 -4.812639 -1e+06 -2.917137 -1e+06 -1e+06

------------------------- level2 -------------------------
                                       1 -> 2        2 -> 2
I((level == "2") * 1)                -1.0317888207  2.7318502453
I((level == "2") * shrub)            -0.0082644265  0.0221857338
I((level == "2") * time_since_release) -0.0002807144 -0.0009640931

                                       3 -> 4 4 -> 4
I((level == "2") * 1)                -1e+06 -1e+06
I((level == "2") * shrub)            -1e+06 -1e+06
I((level == "2") * time_since_release) -1e+06 -1e+06

                                       5 -> 6 6 -> 6
I((level == "2") * 1)                -1e+06 -1e+06
I((level == "2") * shrub)            -1e+06 -1e+06
I((level == "2") * time_since_release) -1e+06 -1e+06

-------------------------------------------------------------

-------------------------------------------------------------
Transition probability matrix (based on mean covariate values):
-------------------------------------------------------------
------------------------- level1 -------------------------
            nonInfected  Infected      Death
nonInfected  0.9405791 0.05177737 0.007643511
Infected     0.0000000 0.94868713 0.051312872
Death        0.0000000 0.00000000 1.000000000
```

```
------------------------- level2 --------------------------
              e          r
e 0.75698486 0.2430151
r 0.06095609 0.9390439

   i id
i  1  0
id 1  0

   d dd
d  1  0
dd 1  0

-------------------------------------------------------------

-------------------------------------------------
Regression coeffs for the initial distribution:
-------------------------------------------------
-------------------- level1 --------------------
             state 3   state 5
(Intercept) -52.42732 -71.61341

-------------------- level2 --------------------
                            state 2
I((level == "2i") * 1) 1.669756

                            state 4
I((level == "2i") * 1) 2.959402e-09

                            state 6
I((level == "2i") * 1) 2.959383e-09

-------------------------------------------------
```

## Section S5. 5-state HMMs

## 5.1. Unconstrained 5-state model

```
Value of the maximum log-likelihood: -822342.8

step parameters:
----------------
          SE        SR       IE       IR        D
mean 1959.649 383.6203 21.26635 2.658252 2.1129006
sd   1052.144 372.7165 18.93075 1.490357 0.5262408

angle parameters:
-----------------
                      SE         SR         IE        IR         D
mean          -0.03113489 0.04102629 -3.1293622 3.1312131 2.631227
concentration  1.56198632 0.16905466  0.3584423 0.3339477 6.648088

Regression coeffs for the transition probabilities:
---------------------------------------------------
                           1 -> 2         1 -> 3    1 -> 4    1 -> 5       2 ->
1       2 -> 3      2 -> 4    2 -> 5         3 -> 1
(Intercept)      -1.0162908447 -2.685773e+00 -2.990627 -1.834031
-2.878513e+00 -2.187746441 -2.3379370164 -24.96851 -3.226942e+00
shrub            -0.0219192633  1.976282e-03 -6.022947 -2.349271
-2.943447e-02 -0.006404594  0.0247462881   0.00000 -2.812501e-02
time_since_release -0.0004811317 -1.513205e-05 -1.164282 -1.349730
-3.232238e-05  0.000378894 -0.0008349726   0.00000  8.185061e-05
                           3 -> 2         3 -> 4    3 -> 5       4 -> 1        4
-> 2       4 -> 3    4 -> 5 5 -> 1 5 -> 2 5 -> 3 5 -> 4
(Intercept)      -0.1205061713 -1.357008588 -1.2829509 -3.010103382
-0.0771407536 -1.5728257701 -1.786558 -1e+06 -1e+06 -1e+06 -1e+06
shrub            -0.0307891633  0.007738664 -0.5289330 -0.039744950
-0.0278891124 -0.0249301076 -2.123231 -1e+06 -1e+06 -1e+06 -1e+06
time_since_release -0.0003757599 -0.001686547 -0.6775713 -0.001417873
-0.0002530488  0.0005630142 -1.397285 -1e+06 -1e+06 -1e+06 -1e+06

Transition probability matrix (based on mean covariate values):
---------------------------------------------------------------
           SE         SR         IE           IR            D
SE 0.73786075 0.2112493 0.05088998 1.529051e-102 1.206953e-103
SR 0.03596572 0.7937169 0.09041440  7.990293e-02  1.137577e-11
IE 0.01722097 0.3505810 0.52369601  1.085021e-01  9.591827e-51
IR 0.01514746 0.3738021 0.09761470  5.134357e-01 2.042947e-106
D  0.00000000 0.0000000 0.00000000  0.000000e+00  1.000000e+00

Initial distribution:
---------------------
          SE           SR           IE           IR            D
6.528100e-04 1.267492e-04 6.851729e-01 3.140378e-01 9.794411e-06
```

## 5.2. Constrained 5-state model

```
Value of the maximum log-likelihood: -829127.5

step parameters:
----------------
          SE       SR       IE       IR        D
mean 1891.260 296.2557 379.8396 4.315509 5.820463
sd   1185.073 414.9153 454.7442 3.261060 5.241028

angle parameters:
-----------------
                     SE          SR         IE        IR         D
mean        -0.0334418 1.823797672 0.02155107 3.1385940 3.1092579
concentration 1.8121634 0.008826148 0.13890107 0.4392679 0.8135199

Regression coeffs for the transition probabilities:
---------------------------------------------------
                          1 -> 2 1 -> 3 1 -> 4 1 -> 5        2 -> 1 2 -> 3        2
-> 4  2 -> 5 3 -> 1 3 -> 2        3 -> 4 3 -> 5 4 -> 1
(Intercept)        -9.658813e-01 -1e+06 -1e+06 -1e+06 -2.3667863026 -1e+06
-6.6977437749 -14.263 -1e+06 -1e+06 -2.3143799222 -1e+06 -1e+06
shrub               -2.603151e-02 -1e+06 -1e+06 -1e+06  0.0024873074 -1e+06
0.1120207665   0.000 -1e+06 -1e+06  0.0294247581 -1e+06 -1e+06
time_since_release -6.494087e-05 -1e+06 -1e+06 -1e+06  0.0005266694 -1e+06
0.0002775426   0.000 -1e+06 -1e+06 -0.0003619823 -1e+06 -1e+06
                          4 -> 2        4 -> 3    4 -> 5 5 -> 1 5 -> 2 5 -> 3 5 -> 4
(Intercept)        -1e+06 -0.1603475754 -6.231782 -1e+06 -1e+06 -1e+06 -1e+06
shrub              -1e+06 -0.0264425082  0.000000 -1e+06 -1e+06 -1e+06 -1e+06
time_since_release -1e+06 -0.0002612298  0.000000 -1e+06 -1e+06 -1e+06 -1e+06

Transition probability matrix (based on mean covariate values):
---------------------------------------------------------------
           SE         SR        IE          IR            D
SE 0.76178139 0.2382186 0.0000000 0.000000000 0.000000e+00
SR 0.09391075 0.9034921 0.0000000 0.002596575 5.775399e-07
IE 0.00000000 0.0000000 0.8968311 0.103168870 0.000000e+00
IR 0.00000000 0.0000000 0.4028632 0.595965214 1.171635e-03
D  0.00000000 0.0000000 0.0000000 0.000000000 1.000000e+00

Initial distribution:
---------------------
          SE          SR           IE          IR            D
0.0092222272 0.7275775009 0.0263061181 0.2367375727 0.0001565811
```

## 5.3 Constrained 5-state model with known death <span style="color:red">(no convergence)</span>

```
Value of the maximum log-likelihood: -1.797693e+308

step parameters:
----------------
       SE  SR  IE IR   D
mean 1800 750 220 15 2.5
sd   1050 484 220 12 1.0

angle parameters:
-----------------
```

```
                    SE     SR    IE    IR      D
mean             -0.03 -0.009  0.40  -3.1 3.141593
concentration     1.80  0.530  0.01   0.3 0.300000

Regression coeffs for the transition probabilities:
---------------------------------------------------
                  1 -> 2 1 -> 3 1 -> 4 1 -> 5 2 -> 1 2 -> 3 2 -> 4 2 -> 5 3 -> 1 3
-> 2 3 -> 4 3 -> 5 4 -> 1 4 -> 2 4 -> 3 4 -> 5 5 -> 1
(Intercept)        -1.5 -1e+06 -1e+06 -1e+06   -1.5   -1.5 -1e+06
-1e+06   -1.5 -1e+06 -1e+06 -1e+06   -1.5   -1.5 -1e+06
shrub               0.0 -1e+06 -1e+06 -1e+06    0.0    0.0 -1e+06
-1e+06    0.0 -1e+06 -1e+06 -1e+06    0.0    0.0 -1e+06
time_since_release  0.0 -1e+06 -1e+06 -1e+06    0.0    0.0 -1e+06
-1e+06    0.0 -1e+06 -1e+06 -1e+06    0.0    0.0 -1e+06
                  5 -> 2 5 -> 3 5 -> 4
(Intercept)        -1e+06 -1e+06 -1e+06
shrub              -1e+06 -1e+06 -1e+06
time_since_release -1e+06 -1e+06 -1e+06

Transition probability matrix (based on mean covariate values):
--------------------------------------------------------------
          SE        SR        IE        IR        D
SE 0.8175745 0.1824255 0.0000000 0.0000000 0.0000000
SR 0.1336597 0.5990210 0.0000000 0.1336597 0.1336597
IE 0.0000000 0.0000000 0.8175745 0.1824255 0.0000000
IR 0.0000000 0.0000000 0.1542808 0.6914385 0.1542808
D  0.0000000 0.0000000 0.0000000 0.0000000 1.0000000

Initial distribution:
---------------------
 SE SR IE IR  D
0.2 0.2 0.2 0.2 0.2
```

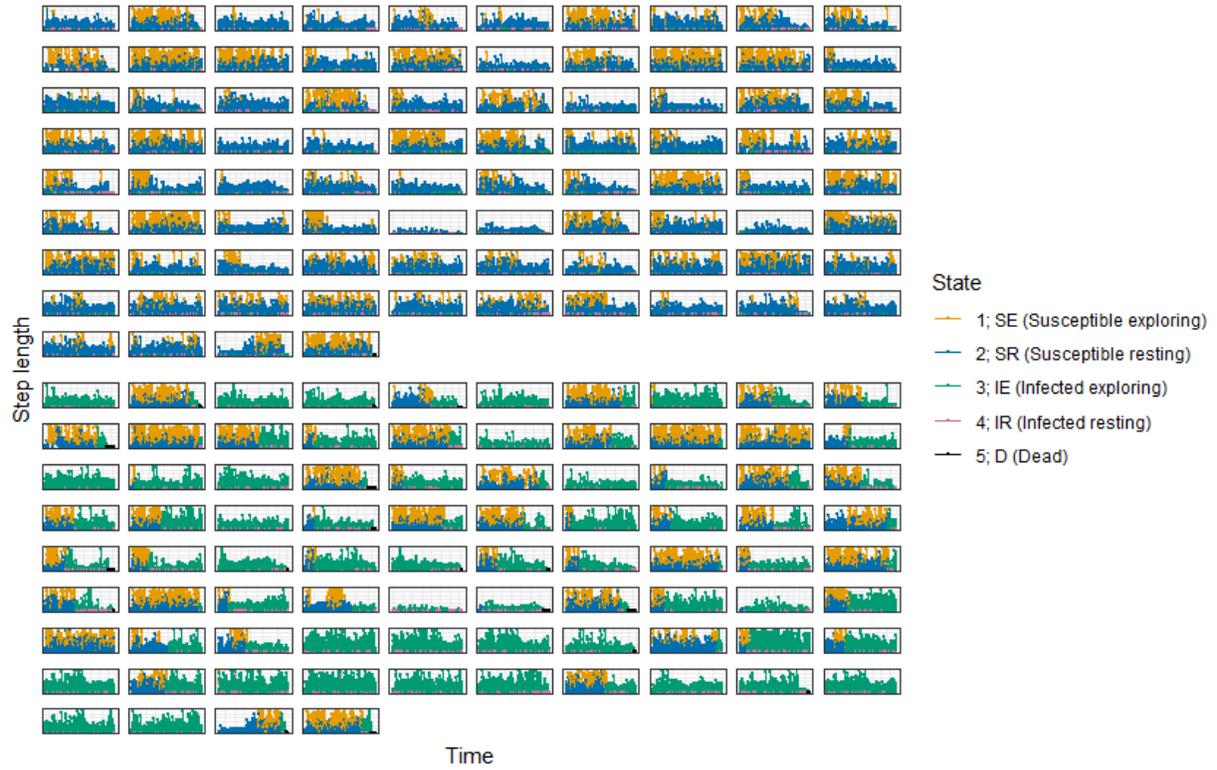

**Figure S7.** A time series of step lengths colored by the decoded states from the unconstrained 5-state HMM (a top panel) vs the constrained 5-state HMM (a bottom panel)

| | | HMM | | HHMM | |
|---|---|---|---|---|---|
| | **State** | **unconstrained** | **constrained** | **unconstrained** | **constrained** |
| Step length (mean) | SE | 1803 | 1629 | 1691 | 1676 |
| | SR | 318 | 275 | 298 | 284 |
| | I | 4 | 194 | 202 | 217 |
| | D | 8 | 14 | 61 | 65 |
| Turning angle (mean/concentration) | SE | -0.04 / 1.5 | -0.04 / 1.6 | -0.04 / 1.58 | -0.03 / 1.5 |
| | SR | 0.08 / 0.08 | $\pi$ / 0.03 | $\pi$ / 0 | 2.4 / 0 |
| | I | $\pi$/0.4 | 2.79 / 0.05 | -$\pi$ / 0.07 | -$\pi$ / 0.04 |
| | D | $\pi$/0.77 | -$\pi$ / 0.59 | $\pi$ / 0.26 | $\pi$ / 0.26 |

**Table S3.** A summary of the estimated mean step length and turn angle distributions for each state (Susceptible Exploring; SE, Susceptible Resting; SR, Infected; I, and Dead; D) from the models fitted to the oryx data. Note that the interpretations of the states are tentative and, based on our results, they are not appropriate for the unconstrained HMM.

| **State** | **accuracy** | **precision** | **recall** | **F1** |
|---|---|---|---|---|
| S | 0.998 | 0.992 | 1 | 0.996 |
| I | 0.994 | 0.986 | 0.989 | 0.987 |
| R | 0.999 | 0.998 | 1 | 0.999 |
| D | 0.997 | 1.000 | 0.982 | 0.991 |

**Table S4.** A summary of the constrained SIR HMM's performance metrics for each state (Susceptible; S, Infected; I, Recovered; R, and Dead; D) that compares the true states from the simulation and estimates the state from the model.

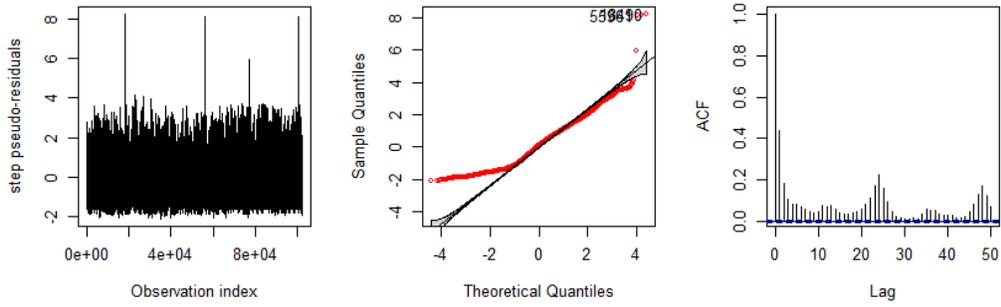

**Figure S8.** Pseudo-residual plot for the constrained 3-state HMM.

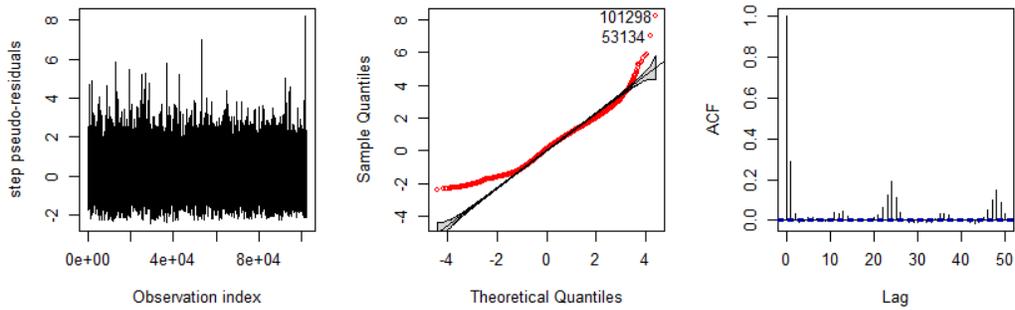

**Figure S9.** Pseudo-residual plot for the constrained 4-state HMM.

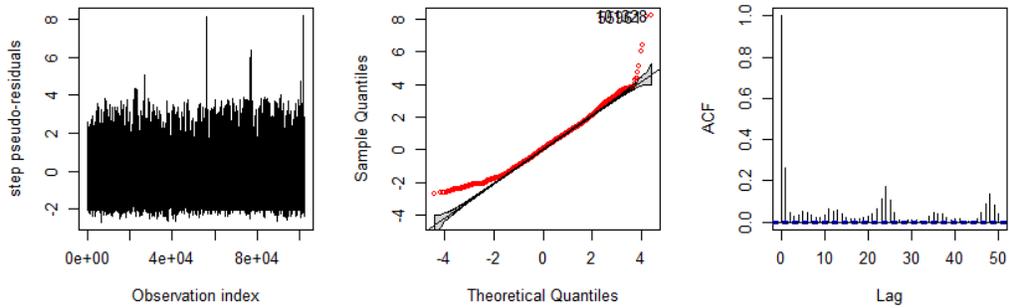

**Figure S10.** Pseudo-residual plot for the constrained 5-state HMM.

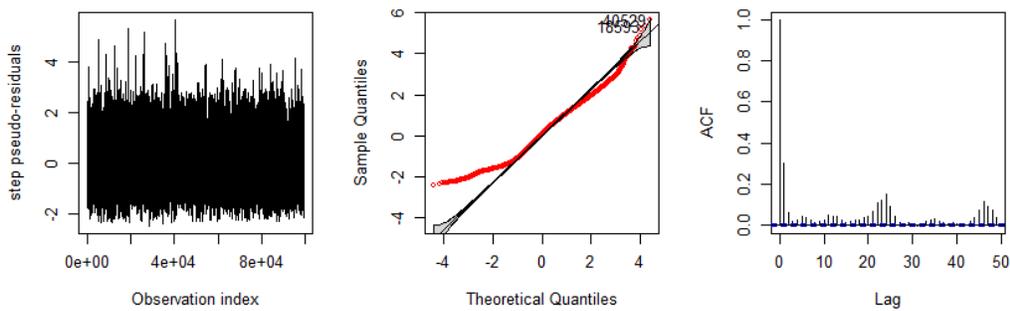

**Figure S11.** Pseudo-residual plot for the constrained 4-state HHMM.